\documentclass{article}
\usepackage{RR}

\usepackage{epsf}
\usepackage{amsmath}
\usepackage{latexsym}
\usepackage{amssymb}
\usepackage{amsthm}
\usepackage{times}
\usepackage{rotating}
\usepackage{setspace}
\usepackage{graphicx}
\usepackage{multicol}
\usepackage{latexsym}
\usepackage{amsfonts}
\usepackage{subfigure}

\newtheorem{theorem}{Theorem}[section]
\newtheorem{lemma}[theorem]{Lemma}

\newcounter{linecounter}
\newcommand{\linenumbering}{(\arabic{linecounter})}
\renewcommand{\line}[1]{\refstepcounter{linecounter}
\label{#1}
\linenumbering}
\newcommand{\resetline}{\setcounter{linecounter}{0}}
\newif\ifcode
\codefalse

\usepackage{macros}
\ifcode
\usepackage[ruled]{algorithm}
\usepackage{ioa_code}
\fi

\newcommand{\remove}[1]{}

\begin{document}

\RRdate{D\'ecembre 2006}
%\RRNo{ }

\title{{\bf Distributed Slicing in Dynamic Systems}}

\RRauthor{{\bf Antonio Fern\'andez\thanks{Universidad Rey Juan Carlos, 28933 M\'ostoles, Spain. anto@gsyc.escet.urjc.es}\and Vincent Gramoli\thanks[2]{IRISA, INRIA Universit\'e Rennes 1 (ASAP Research Group) 35042 Rennes, France. \{vgramoli,akermarr,raynal\}@irisa.fr}\and Ernesto Jim\'enez\thanks{Universidad Polit\'ecnica de Madrid, 28031 Madrid, Spain. ernes@eui.upm.es}\and \\
Anne-Marie Kermarrec\thanksref{2}\and 
Michel Raynal\thanksref{2}}
}
\authorhead{Fern\'andez \& Gramoli \& Jimenez \& Kermarrec \& Raynal} %% Ceci apparait sur chaque page paire.
\RRetitle{Distributed Slicing in Dynamic Systems} %% english title
\titlehead{Distributed Slicing in Dynamic Systems} %%titre court, sur chaque page impaire.
\RRtitle{Morcellement distribu\'e dans les syst\`emes dynamiques} 
\RRnote{Contact author: Vincent Gramoli \texttt{vgramoli@irisa.fr}}

\RRabstract{Peer to peer (P2P) systems are moving from application specific architectures to 
a generic service oriented design philosophy.
This raises interesting problems in connection with providing useful
P2P middleware services that are capable of dealing with resource assignment
and management in a large-scale, heterogeneous and unreliable environment.
One such service, the slicing service, has been proposed to allow for an
automatic partitioning of P2P networks into groups (slices) that
represent a controllable amount of some resource and that are also relatively
homogeneous with respect to that resource, in the face of churn and
other failures.
In this report we propose two algorithms to solve the distributed slicing problem.
The first algorithm improves upon an existing algorithm that is based on 
gossip-based sorting of a set of uniform random numbers.
We speed up convergence via a heuristic for gossip peer selection.
The second algorithm is based on a different approach: statistical approximation
of the rank of nodes in the ordering.
The scalability, efficiency  and resilience to dynamics of both algorithms 
relies on their gossip-based models.
We present theoretical and experimental results to prove the viability
of these algorithms.}

\RRresume{Un service de \emph{morcellement} d'un r\'eseau pair-\`a-pair permet de partitionner les n\oe uds du syst\`eme en plusieurs groupes appel\'es \emph{morceaux}.  Ce rapport pr\'esente deux algorithmes pour r\'esoudre le probl\`eme du morcellement r\'eparti.  Le premier algorithme am\'eliore un algorithme existant en acc\'erant son temps de convergence.
Le second algorithme utilise une approche diff\'erente d'appoximation statistique. Des r\'esultats th\'eoriques et experimentaux
montrent la viabilit\'e de nos algorithmes.}

\RRmotcle{Morcellement, Bavardage, Morceau, Va-et-vient, Pair-\`a-pair, Aggr\'egation, Grande \'echelle, Allocation de ressources.}
\RRkeyword{Slicing, Gossip, Slice, Churn, Peer-to-Peer, Aggregation, Large Scale, Resource Allocation.}
%% \RRprojet{Bibli} % cas d'un seul projet

%\RRprojets{Bibli et Ami} %% cas de 2 projets.
\RRprojet{Asap}
%% \RRtheme{\THBio} % cas d'un seul theme
\RRtheme{\THCom} %% cas de 2 themes

\URRennes

\makeRR
%
%\author{
%\authorblockN{Antonio Fern\'andez\authorrefmark{1},
%Vincent Gramoli\authorrefmark{2}, 
%%Mark Jelasity\authorrefmark{3}, 
%Ernesto Jimenez\authorrefmark{4},  \\
%Anne-Marie Kermarrec\authorrefmark{2}, 
%Michel Raynal\authorrefmark{2}
%} \\
%\authorblockA{
%\hspace{-0.5cm}\begin{tabular}{ccccccc}
%\authorrefmark{1}{\small Universidad Rey} &  & \authorrefmark{2}{\small IRISA, INRIA}  & & \authorrefmark{4}{\small Universidad Polit\'ecnica} \\
%{\small Juan Carlos,} & & {\small et Universit\'e Rennes 1,}    & & {\small de Madrid,} \\
%{\small 28933 M\'ostoles, Spain.} & & {\small 35042 Rennes, France.}    & & {\small 28031 Madrid, Spain.} \\ 
%{\small anto@gsyc.escet.urjc.es} & &  {\small \{vgramoli,akermarr,raynal\}@irisa.fr} & & {\small ernes@eui.upm.es}
%\end{tabular}}}

%\newcommand{\pb}{\emph Self-Organizing Locality-based Algorithm for Distributed Slicing}

\maketitle

%\doublespace

%\begin{abstract}
%Peer to peer (P2P) systems are moving from application specific architectures to 
%a generic service oriented design philosophy.
%This raises interesting problems in connection with providing useful
%P2P middleware services that are capable of dealing with resource assignment
%and management in a large-scale, heterogeneous and unreliable environment.
%One such service, the slicing service, has been proposed to allow for an
%automatic partitioning of P2P networks into groups (slices) that
%represent a controllable amount of some resource and that are also relatively
%homogeneous with respect to that resource, in the face of churn and
%other failures.
%In this report we propose two algorithms to solve the distributed slicing problem.
%The first algorithm improves upon an existing algorithm that is based on 
%gossip-based sorting of a set of uniform random numbers.
%We speed up convergence via a heuristic for gossip peer selection.
%The second algorithm is based on a different approach: statistical approximation
%of the rank of nodes in the ordering.
%The scalability, efficiency  and resilience to dynamics of both algorithms 
%relies on their gossip-based models.
%We present theoretical and experimental results to prove the viability
%of these algorithms. \\
%\end{abstract}
%\vspace{-0.3cm}
%\noindent
%Keywords: Slicing, Gossip, Slice, Churn, Peer-to-Peer, Aggregation, Large Scale, Resource Allocation. \\
%Technical Areas: Operating Systems and Middleware, Internet Computing and Applications, Peer-to-Peer. \\
%Contact Author: Vincent Gramoli, vgramoli@irisa.fr

\section{Introduction}

\subsection{Context and Motivations}

The peer to peer (P2P) communication paradigm has now become the prevalent
model to build large-scale distributed applications, able to cope  with both 
scalability and system dynamics.  This is now a mature technology:
peer to peer systems are slowly moving from application-specific
 architectures to a generic-service oriented design philosophy. 
More specifically, peer to peer protocols hold the promise to integrate into
platforms on top of which several applications with various requirements
 may cohabit. This leads to the interesting issue of resource 
assignment or how to allocate a set of nodes for a given application. 
Examples of  targeted platforms for such a service are telecommunication 
platforms, where some set of peers may be automatically assigned to a 
specific task depending on their capabilities,
testbed platform such as Planetlab~\cite{BBC+04}, 
or desktop-grid-like applications~\cite{A04}.
 
In this context, the ordered slicing service has been 
recently proposed as a building block to allocate resources,
i.e a set of nodes sharing some characteristics with respect to  a
given metric or attribute,  in a large-scale peer to peer
system. This service acknowledges the fact that peers 
potentially offer heterogeneous capabilities as 
revealed by many recent works describing heavy-tailed distributions
 of storage space, bandwidth, and uptime of peers~\cite{SGG02,BSV03,SR06}.
 The slicing  service \cite{JK06} enables peers in a large-scale
 unstructured network to self-organize into a partitioning,
 where partitions (slices) are connected overlay networks that represent
 a given percentage of some resource.
Such slices
can be allocated to specific applications later on.
The slicing is ordered in the sense that peers get ranked
 according to their capabilities expressed by an attribute value.

Large scale dynamic distributed systems consist of many 
participants that can join and leave 
at will. Identifying peers in such systems that have a similar level
 of power or capability (for instance, in terms of bandwidth, 
processing power, storage space, or uptime) in a completely 
decentralized manner is a difficult task. It is even harder 
to maintain this information in the presence of churn. 
 Due to the intrinsic dynamics of contemporary peer to peer
 systems it is impossible to obtain accurate information about 
the capabilities (or even the identity) of the system participants. 
Consequently, no node is able to maintain accurate information about 
all the nodes. This  disqualifies  centralized approaches.  

Taking this into account, we can summarize the ordered slicing problem
we tackle in this report:
we need to rank nodes depending on their capability, slice the network 
depending on these capabilities and, most importantly, readapting
the slices continuously to cope with system dynamism.  

Building  upon the work on ordered distributed slicing
%%MJ I removed "seminal"; felt it's too self-promoting
proposed in \cite{JK06}, here we focus on the issue of \textit{accurate}
 slicing.
That is, we focus on improving the quality and stability of the slices, both
aspects being crucial for potential applications.

%A slice is 
%composed of a subset of nodes which are attribute values close 
%to each other. Additionally we can consider slices ordered, 
%so that the first slice contains the nodes with the
% smallest attribute values while the last one contains 
%the nodes with the largest attribute values.
%A node determines the slice it belongs to by comparing
% its attribute value with the values of other nodes.
%Roughly speaking, depending on the attribute values of the other nodes 
%it encounters, a node $i$  determines if its attribute value 
%is relatively high or low and where it lies in the
%set of nodes. Thus, node $i$ tries to estimate 
%its rank (dictated by its capability) to determine 
%the slice it belongs to.
%%
%%MJ I felt the paragraph was reduntant and too detailed for an intro.
%%besides, we don't need it since nothing is built on this info here

\subsection{Contributions}

%%MJ moved para below here from before the section heading
%%MJ also commented it out, becuase it is redundant
%% note that the intro is very long, and this section in particular

%This report presents two main contributions. The first contribution is
%the speedup of the convergence of the \cite{JK06} algorithm. We then 
%identified a few issues of~\cite{JK06} related to slice 
%consistency under churn correlated to attribute values. 
%We fix these issues
% by proposing a second algorithm where each node 
%continuously approximates its rank relatively to the other peers.
% Both algorithms rely on  a gossip-based communication model,
%which has proved to be both lightweight and highly resilient to 
%system dynamics. 

The report presents two gossip-based solutions to slice the nodes
 according to their capability (reflected by an attribute value) 
 in a distributed manner with high probability.
The first contribution of the report builds upon the distributed 
slicing algorithm proposed in \cite{JK06} that we call the JK algorithm in the
sequel of this report.
The second algorithm is a different approach based on rank approximation
through statistical sampling.

In JK, each node $i$  
maintains a random number $r_i$, picked up uniformly at random (between 0 and 1),  
and an attribute value $a_i$, expressing its capability according to a given
metric. 
Each peer periodically gossips with another peer $j$, 
randomly chosen among the peers it knows about. If the 
order between $r_j$ and $r_i$ is different than 
the order between $a_j$ and $a_i$, random values 
are swapped between nodes. The algorithm ensures that 
eventually the order on the 
random values matches the order of the attribute ones.  
The quality of the ranking can then be measured by using a
 global disorder measure expressing
the difference between the exact rank and the actual rank of each peer along the attribute value.

 The first contribution of this report is to propose a local disorder
 measure so that a peer chooses the neighbor to communicate with in order to  maximize
the chance of decreasing the global disorder measure.
 The interest of this approach is to speed the convergence up. 
We provide the analysis and experimental results of this improvement.

Once peers are ordered along the attribute values, the slicing in JK 
takes place as follows. Random values are used
 to calculate which slice a node belongs to. For example,  
a slice containing 20\% of the  best nodes according to a given attribute, 
will be composed of the nodes that end up holding random values greater than 0.8. 
The accuracy of the slicing (independent from the accuracy of the ranking)
 fully depends on the uniformity of the random value
 spread between 0 and 1 and the fact that the proportion of 
random values between 0.8 and 1 is approximately (but usually not exactly)
20\% of the nodes. Another contribution of this report is to precisely 
characterize the potential inaccuracy resulting from this effect.

This observation means that the problem of ordering nodes 
based on uniform random values is not fully sufficient for
determining slices.
This motivates us to find an alternative
approach to this algorithm and JK
in order to determine more precisely the slice each node belongs to.

Another motivation for an alternative approach is related to churn and dynamism.
It may well happen that the
 churn is actually correlated  
to the attribute value. For example, if the peers are 
sorted according to their connectivity potential, 
a portion of the attribute space (and therefore the random value space) might be 
suddenly affected. 
New nodes will then pick up new random values and eventually the distribution of random values will be skewed towards high values.

The second contribution is an alternative algorithm solving these issues  
by approximating the rank of the nodes in the ordering locally, without
the application of random values.
  The basic idea is that each node periodically estimates its rank 
along the attribute axis depending of the attributes it has seen so far.
This algorithm is robust and lightweight due to its
gossip-based communication pattern: each node communicates 
periodically with a
restricted dynamic neighborhood that guarantees connectivity and provides
a continuous stream of new samples.
Based on continuously aggregated information, 
the node can determine the
slice it belongs to with a decreasing error margin.  
 We show that this algorithm provides accurate estimation at the price of a slower convergence.

\subsection{Outline}
The rest of the report is organized as follows: Section \ref{related} surveys some related work. 
The system model is presented in Section \ref{model}. 
The first contribution of an improved ordered slicing algorithm based 
on random values is presented  in Section~\ref{JK+} and the second algorithm 
based on dynamic ranking in Section~\ref{sec:ranking}. 
%Section \ref{sec:dicussion} provide some material for discussion before concluding in Section \ref{conclusion}.
Section~\ref{conclusion} concludes the report.

\section{Related Work}
\label{related}
%A more general problem than the one investigated here 
%is the sorting problem, where each node is ordered among others.
%A version of this problem has been referred as the 
%\emph{external sorting problem}~\cite{DNS91}.  What caught our attention is that 
%this provides a distributed sorting
%algorithm where the memory space of each processor does not necessarily 
%depend on the input and it outputs a sorted sequence of values
%distributed among processors.  
%The \emph{percentile finding} problem was defined in~\cite{IRV89} as
%dividing a set of values into equally sized sets.

Most of the solutions proposed so far for ordering nodes come
from the context of databases,
where parallelizing query executions is used to improve efficiency. 
A large majority of the solutions in this area rely on centralized gathering
or all-to-all exchange, which makes them unsuitable 
for large-scale networks.
For instance, the \emph{external sorting problem}~\cite{DNS91} consists in 
providing
a distributed sorting algorithm where the memory space of each 
processor does not necessarily depend on the input. This algorithm
must output a sorted sequence of values distributed among 
processors.  
The solution proposed in~\cite{DNS91} needs a global merge of 
the whole information, and
thus it implies a centralization of information.
Similarly, the \emph{percentile finding} problem~\cite{IRV89}, which 
aims at dividing a set of values into equally sized sets,
requires a logarithmic number of all-to-all message exchanges.
%Despite the unaffordable number of messages it requires, it would 
%be worthy to investigate an adaptation of this algorithm using
%gossip-based algorithm for the slice partitioning problem.

Other related problems are the selection problem and the $\phi$-quantile search.
The selection problem~\cite{FR75,BFPRT72} aims at determining the $i^{th}$
smallest element with as few comparisons as possible.
The $\phi$-\emph{quantile} search (with 
$\phi \in (0,1]$) is the problem to find among $n$ elements the $(\phi n)^{th}$ element.
Even though these problems look similar to our problem, 
%in the sense
%that they all focus on finding the ordering of a node among the system, it remains 
%different in the following sense: 
they aim 
at finding a specific node among all, while the distributed slicing problem aims
at solving a global problem where each node maintains a piece of
information.  
Additionally, solutions to the quantile search problem like the one presented 
in~\cite{KDG03} use an approximation of the system size. The same holds 
for the algorithm in~\cite{SDCM06}, which uses similar ideas to determine the distribution
of a utility in order to isolate peers with high capability---i.e., super-peers.

%As far as we know, the most similar work appeared in
As far as we know, the distributed slicing problem was studied in a P2P system 
for the first time in~\cite{JK06}. In this report, a node with the $k^{th}$
smallest attribute value, among those in a system of size $n$, tries to estimate
its normalized index $k/n$.
%where 
%each node $i$ is provided with a random value which is an estimate of the 
%rank of node $i$.   
%periodically exchange random value to reflect the rank
%of its attribute value among all attribute values of the system.  
%
The \emph{JK algorithm} proposed in~\cite{JK06} works as follows.
Initially, each node draws independently and uniformly a random value in the
interval $(0,1]$ which serves as its first estimate of its normalized index.
Then, the nodes use a variant of Newscast~\cite{JMB05} to gossip among each
other to exchange random values when they
find that the relative order of their random values and that of their attribute
values do not match.
%The algorithm aims at exchanging random 
%values among peers to reflect the order given by their attribute values.
%When for any $k$ the node with the $k^{th}$ attribute value has also 
%the $k^{th}$ random value, the system is ordered and random value.
%%
This algorithm is robust in face of frequent 
dynamics and guarantees a fast convergence to the same sequence of peers with
respect to the random and the attribute values. At every point in time the
current random value of a node serves to estimate the slice to which it
belongs (its slice).
%Consequently this solution 
%is well-suited for ordering nodes in a large-scale dynamic systems.
%The first algorithm we propose speeds up this decrease.

%The more similar work has been proposed in~\cite{JK06}, the authors investigate 
%the slice partitioning problem in a deterministic way.
%However, their approach requires that each node generates a random 
%value.  More precisely, each node maintains a view containing neighbors 
%information such as their id, their attribute value, and their random value.  
%The algorithm presented in~\cite{JK06} makes a node to select an arbitrary 
%misplaced neighbor to swap its random value with.  A node estimates the slice
%it belongs to according to the rank given by the current random value it owns.
%This solves the slice partitioning problem if the random values are uniformly 
%spread among all possible values.

\section{Model}
\label{model}

\subsection{System model}

We  consider a  system $\Sigma$ containing  a set
 of $n$ uniquely identified nodes. (Value $n$ may vary over time, dynamics is explained below).
%%MJ don't know what this means, but not needed anyway: commented out
% we should use no footnotes at all in any case, or only if it is absolutely
%essential
 %\footnote{The value $n$ is observed instantaneously but may vary over time.}  
The set of identifiers is denoted by $I$.
%and the number of nodes  by $n$.
Each node can leave and new nodes can join the system at any  
time, thus the number of nodes is a function of time.
Nodes may also crash. In this report,
we do not differentiate between a crash and a voluntary node departure. 

Each node $i$ maintains an attribute value $a_i$, reflecting the node
capability according to a specific metric.
These attribute values over the network might have an arbitrary
skewed distribution.
Initially, a node
has no global information neither about the structure or size of the system nor
about 
the attribute values of the other nodes. 

We can define a total ordering over the nodes based on their attribute value,
with the node identifier used to break ties.
Formally, we let $i$ precede $j$ if and only if $a_i < a_j$, or
$a_i = a_j$ and $i < j$. We refer to this totally ordered sequence as the
\emph{attribute-based sequence}, denoted by $A.\ms{sequence}$. The attribute-based
rank of a node $i$, denoted by $\alpha_i \in \{1,...,n\}$, is defined as the index
of $a_i$ in $A.\ms{sequence}$. 
For instance, let us consider three nodes: 1, 2, and 3, with three
different attribute values $a_1 = 50$, $a_2 = 120$, and $a_3 = 25$.
In this case, the attribute-based rank of node $1$ would be
$\alpha_1 = 2$.
In the rest of the report, we assume that nodes are sorted according to a single 
attribute and that each node belongs to a unique slice.
The sorting along several attributes is out of the scope of this report.

%Initially, node $i$ does not know the value of $\alpha_i$.
%%MJ Nor does it learn it later, at least in the ordering approach
%AMK: I don;t think we should mention this indeed.

\subsection{Distributed Slicing Problem}\label{ssec:pb}

Let ${\cal S}_{l,u}$ denote the \emph{slice} containing every node $i$ whose
normalized rank, namely $\frac{\alpha_{i}}{n}$, satisfies $l < \frac{\alpha_{i}}{n}
\leq u$
where $l\in [0,1)$ is the slice lower boundary and $u\in (0,1]$ is 
the slice upper boundary so that all slices represent adjacent intervals $(l_1, u_1], (l_2, u_2]$...
Let us assume that we partition the interval $(0,1]$ using a set of slices,
and this partitioning is known by all nodes.
The distributed slicing problem requires each node 
to determine the slice it currently belongs to.
Note that the problem stated this way is similar to the
ordering problem, where each node has to determine its own index in
$A.\ms{sequence}$.
However, the reference to slices introduces special requirements related to
stability and fault tolerance, besides, it allows for future generalizations
when one considers different types of categorizations.

\begin{figure}
\centering\includegraphics[scale=0.6]{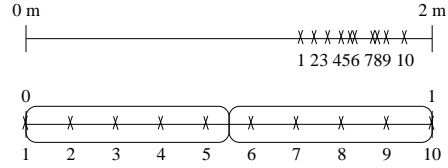}
\caption{Slicing of a population based on a height attribute.}
\label{fig:size}
\end{figure}

%\subsection{Example of Slicing a Population}
 Figure~\ref{fig:size} illustrates an example 
of  a population of 10 persons, to be sorted 
against their height.
A partition of this population could be defined by two slices of the same
size: the group of short persons, and the 
group of tall persons. This is clearly an example 
where the distribution of attribute values is skewed towards 2 meters.
 The rank of each person in
the population and the two slices are represented on the bottom axis. Each
person is represented as a small cross on these axes.\footnote{Note that the shortest (resp. largest) rank is represented by a cross at the 
extreme left (resp. right) of the bottom axis.} 
Each slice is represented as an oval.  The slice $S_1 = {\cal
S}_{0,\frac{1}{2}}$ contains the five shortest persons and 
the slice $S_2 = {\cal S}_{\frac{1}{2},1}$ contains the five tallest
persons.

Observe that another way of partitioning the population could be to define the
group of short persons as that containing all the persons shorter than a 
predefined measure (e.g., $1.65m$) and the group of tall persons as that containing
the persons taller than this measure.
%The groups differs from the slices because the slices rely tightly on 
%the distribution of attribute values:  while one of these three
%groups could be empty even though the population is large, slices have a minimal size
%depending on the population size.
However, this way of partitioning would most certainly lead to an unbalanced
distribution of persons, in which, for instance a group might be empty (while a
slice is almost surely non-empty). Since the distribution of attribute values is
unknown and hard to predict, defining relevant groups is a difficult task.  For
example, if the distribution of the human heights were unknown, then the persons
taller than $1m$ could be considered as tall and the persons shorter than $1m$
could be considered as short.  Conversely, slices partition the population  
into subsets representing a predefined portion of this population.
Therefore, in the rest of the report, we consider slices as defined as a proportion of the
network.

\subsection{Facing Churn}

%%MJ rewrote it from skretch. Still, I'm a bit hesitant, becase we do not
% actually solve these problems, do we?
%AMK: we do somehow in the second case

Node churn, that is, the continuous arrival and departure of nodes is an intrinsic characteristic 
of peer to peer systems and may significantly impact  the outcome, and more specifically the accuracy 
of the slicing algorithm.
The easier case is when  the distribution of the attribute values of the departing and
arriving nodes are identical.
In this case, in principle, the arriving nodes must find their slices, but
the nodes that stay in the system  are mostly able to keep their slice assignment.
Even in this case however, nodes that are close to the border of a slice
may expect frequent changes in their slice due to the variance of the
attribute values, which is non-zero for any non-constant distribution.
If the arriving and departing nodes have different attribute distributions,
 so that the distribution in the actual network of live nodes keeps changing,
then this effect is amplified. However, we believe that this is a realistic assumption
to consider that the churn may be correlated to some specific values (for example 
if the considered attribute is uptime or connectivity).

\section{Dynamic Ordering by Exchange of Random Values}\label{sec:ordering}
\label{JK+}

This section proposes an  algorithm for the distributed slicing 
problem improving upon the original JK algorithm \cite{JK06},
 by considering a local measure of the global disorder function.
In this section we present the algorithm along with the corresponding analysis and 
simulation results.

\subsection{On Using Random Numbers to Sort Nodes}

This Section presents the algorithm built upon JK. 
We refer to this algorithm as \emph{mod-JK} (standing for modified JK).
In JK,
%We build upon  the JK algorithm, where
each node $i$
 generates a number $r_i\in (0,1]$ independently and uniformly at random.
The key idea is to sort these random numbers with 
respect to the attribute values by swapping these random numbers between nodes,
so that if $a_{i} < a_{j}$ then $r_{i} < r_{j}$. 
Eventually,  the attribute values (that are fixed) and the 
random values (that are exchanged)  should be sorted in the same order.
That is, each node would like to obtain the $x^{th}$ largest random number if 
it owns the $x^{th}$ largest attribute value.
Let $R.\ms{sequence}$ denote the \emph{random sequence} obtained 
by ordering all nodes according to their random number.
Let $\rho_i(t)$ denote the index of node $i$ in $R.\ms{sequence}$ at time
$t$. When not required, the time parameter is omitted.

To illustrate the above ideas, consider that nodes 1, 2, and 3 
%from the example above 
from the previous example
have three
distinct random values: $r_1 = 0.85$, $r_2 = 0.1$, and $r_3 = 0.35$.
In this case, the index $\rho_1$ of node $1$ would be $3$. Since the
attribute values are $a_{i}=50$, $a_{2}=120$, and $a_{3}=25$, the algorithm
must achieve the following final assignment of random numbers: $r_{1}=0.35$,
$r_{2}=0.85$, and $r_{3}=0.1$.

Once sorted, the random values are used to determine the portion of the network a peer belongs to.

\subsection{Definitions}

\paragraph{View.}
Every node $i$ keeps track of some neighbors and their age. 
The \emph{age} of neighbor $j$ is a timestamp, $t_j$, set to 0 when $j$ becomes a 
neighbor of $i$.
Thus, node $i$ maintains an array
containing the id, the age,
the attribute value, and the 
random value of its neighbors.  This array, denoted ${\mathcal N}_i$, is 
called the \emph{view} of node $i$. The views of all nodes have the same size, 
denoted by $c$.  
%Neighborhoods are updated regularly following a simple protocol called 
%newscast~\cite{JMB05}.

\paragraph{Misplacement.}
A node participates in the algorithm by exchanging its rank with a misplaced 
neighbor in its view.  Neighbor $j$ is misplaced 
if and only if
\begin{itemize}
\item $a_i >  a_j$ and  $r_i <  r_j$, or
\item $a_i <  a_j$ and  $r_i >  r_j$.
\end{itemize}
We can characterize these two cases by the predicate $(a_j - a_i)(r_j - r_i)<0$.
%We call the \emph{misplaced measurement} of node $i$ 
%the difference $|\frac{a_i}{n} - r_i|$, and the larger this value is,
%the more $i$ is misplaced  (if this value is large, then $i$ estimate is far 
%from reflecting its real rank).
%%MJ you never actually use this definition

\paragraph{Global Disorder Measure.}

In~\cite{JK06}, a measure of the relative
disorder of sequence $R.\ms{sequence}$ with respect to sequence
$A.\ms{sequence}$ was introduced, called the 
\emph{global disorder measure (GDM)} and defined, for any time $t$, as
$$\ms{GDM}(t) = \frac{1}{n}\sum_{i}(\alpha_i - \rho(t)_i)^2.$$

The minimal value of GDM is 0, which is obtained
when $\rho(t)_i = \alpha_i$ for all nodes $i$.  In this case the 
attribute-based index of a node is equal to its random value index, indicating
that random values are ordered.

\subsection{Improved Ordering Algorithm}

In this algorithm, each node $i$ searches its own view ${\mathcal N}_i$
 for misplaced neighbors. Then, 
one of them is chosen to swap  random value with. This process is repeated
until there is no global disorder.
In this version of the algorithm, we provide each node with  the capability
of measuring locally the disorder. This leads to  a new
heuristic for each node to determine 
the neighbor to exchange with which decreases most the disorder.

The proposed technique attempts to decrease the global
disorder in each exchange as much as possible via selecting the neighbor
from the view that minimizes the local disorder (or, equivalently,
maximizes the order \emph{gain}) as defined below.

For a node $i$ to evaluate the gain of exchanging with a node $j$
of its current view ${\mathcal N}_i$,
we define its \emph{local disorder measure} (abbreviated \emph{LDM}$_i$).  Let $LA.\ms{sequence}_i$ and 
$LR.\ms{sequence}_i$ be the local attribute sequence and the local 
random sequence of node $i$, respectively.   These sequences are computed locally by $i$ 
using the information ${\mathcal N}_i \cup \{i\}$.  Similarly to $A.\ms{sequence}$
and $R.\ms{sequence}$, these are the sequences of neighbors where each node
is ordered according to its attribute value and random number, respectively.
Let, for any $j\in {\cal N}_i \cup \{i\}$, $\ell\rho_j(t)$ and $\ell\alpha_j(t)$ be the 
indices of $r_j$ and $a_j$ in sequences $LR.\ms{sequence}_i$ and
$LA.\ms{sequence}_i$, respectively, at time $(t)$.
At any time $t$, the local disorder measure of node $i$ is defined as:
$$\ms{LDM}_i(t) = \frac{1}{c+1}\sum_{j\in {\mathcal N}_i(t)\cup \{i\}}(\ell\alpha_j(t) - \ell\rho_j(t))^2.$$
We denote by $G_{i,j}(t+1)$ the reduction on this measure that $i$ obtains after
exchanging its random value with node $j$ between 
time $t$ and $t+1$. We define it as:
%{\small
\begin{eqnarray} 
G_{i,j}(t+1) &=& \ms{LDM}_i(t) - \ms{LDM}_i(t+1), \notag \\
G_{i,j}(t+1) &=& \frac{(\ell\alpha_i(t) - \ell\rho_i(t))^2 + (\ell\alpha_j(t) -
\ell\rho_j(t))^2
- (\ell\alpha_i(t) - \ell\rho_j(t))^2 - (\ell\alpha_j(t) - \ell\rho_i(t))^2}{c+1}.
\label{eq:gain} 
\end{eqnarray}
%}

The heuristic used chooses for node $i$ the misplaced neighbor $j$ that
maximizes $G_{i,j}(t+1)$.
%\begin{eqnarray} 
%$G_{i,k}(t+1) &=& \tau_i(t) - \tau_i(t+1), \notag \\
% %The gain that $i$ obtains after swapping with $k$ can be expressed as:
% G_{i,k}(t+1) &=& (\alpha_i - \rho_i)^2 + (\alpha_k - \rho_k)^2 - (\alpha_i - \rho_k)^2 - (\alpha_k - \rho_i)^2. \notag 
% \end{eqnarray}
% That is, the node $k$ that maximizes the gain verifies, for any $j\in V_i$:
% \begin{eqnarray} 
% G_{i,j}(t+1) &\leq& G_{i,k}(t+1), \notag \\
% % (p_j - r_j(t))^2 - (p_i - r_j(t))^2 - (p_j - r_i(t))^2 &\leq& (p_k - r_k(t))^2 - (p_i - r_k(t))^2 - (p_k - r_i(t))^2, \notag \\
% \alpha_i \rho_j(t) + \alpha_j \rho_i(t) - \alpha_j \rho_j(t)  &\leq& \alpha_i \rho_k(t) + \alpha_k \rho_i(t) - \alpha_k \rho_k(t). \Box\notag
%\end{eqnarray}

\subsubsection{Sampling Uniformly at Random}

The algorithm relies on the fact that potential misplaced nodes are found
so that they can swap their
random numbers thereby increasing order. 
If the global disorder is high,  it is very likely that
any given node has misplaced neighbors in its view to exchange with.
%
%Nevertheless, the ordering slows down drastically when only few nodes are misplaced.
%To ensure continuous ordering, the underlying membership protocol must provide
%overlying layer with samples as uniformly drawn as possible.
Nevertheless, as the system gets ordered, it becomes more unlikely for a node
$i$ to have misplaced neighbors.
In this stage the way the view is composed plays a crucial role: if fresh
samples from the network are not available, convergence can be slower than
optimal.
%This problem is exacerbated if the neighbors
%of node $i$ are not drawn uniformly, since then some nodes, misplaced with
%respect to $i$, tend to 
%remain misplaced since they never show up in the view of $i$.
%%MJ This is not true in this way, as non-uniform can even be better than
% uniform. I said something softer.

Several protocols may be used to provide a random and dynamic sampling 
in a peer to peer system such as Newscast~\cite{JMB05}, Cyclon~\cite{VGS05} or Lpbcast~\cite{JGKS04}.
They differ mainly by their \textit{closeness}
to the uniform random sampling of the neighbors and the way they handle churn.
In this report,
we chose to  use a variant of the Cyclon
protocol to construct and update the views~\cite{I05}, as it is reportedly
the best approach to achieve a uniform random neighbor set for all nodes.
%%%%% show that Cyclon creates more random views whereas Newscast
%tends to clusterize the communication graph.  Technically, 
%the Cyclon graph (in which links are defined by the views) has a clustering
%coefficient similar to the 
%clustering coefficient of a random graph~\cite{ER59}.  
%More precisely, the coefficient provided by Cyclon is more than 1800 time closer 
%to the coefficient of a random graph, than the coefficient provided by Newscast.
%
%A direct consequence is that with Cyclon a node has a higher chance of
%discovering new nodes than with Newscast so the algorithm described in
%Figure~\ref{alg:rand}
%can be expected to offer better convergence speed.

%AMK: I actually removed almost evrything here and certainly did not claim 
%that was a  contribution.

\subsubsection{Description of the Algorithm}
\label{sec:modifiedjk}

\begin{table}
  \centering
  \begin{tabular}{|c|l|}
\hline
{\small {\bf Variable}} & {\small {\bf Description}} \\ \hline \hline
$j$ & {\small the identifier of the neighbor} \\ \hline
$t_j$ & {\small the age of the neighbor} \\ \hline
$a_j$ & {\small the attribute value of the neighbor} \\ \hline
$r_j$ & {\small the random value of the neighbor} \\ \hline
\end{tabular}
\caption{The array corresponding to the view entry of the neighbor $j$ ($j\in {\cal N}_i$).}
\label{table:entry}
\end{table}

The algorithm is presented in Figure~\ref{alg:rand}.  The active thread at node $i$
runs the membership (gossiping) procedure ($\lit{recompute-view}()_i$)
and the 
exchange of random values periodically.
As motivated above, the membership procedure, specified in
Figure~\ref{alg:rcyclon}, is similar to the Cyclon algorithm: 
each node $i$ maintains a view ${\mathcal N_i}$ containing one
entry per neighbor.  The entry of a neighbor $j$ corresponds to a tuple
presented in Table~\ref{table:entry}.  Node $i$ copies its view, selects the
oldest neighbor $j$ of its view, removes the entry $e_{j}$ of $j$ from the copy
of its
view, and finally sends the resulting copy to $j$.  When $j$ receives the
view, $j$ sends its own view back to $i$ discarding possible pointers to 
$i$, and $i$ and $j$ update their view with the one they 
receive. This variant of Cyclon, as opposed to the original version,
exchanges all entries of the view at each step.

\begin{figure}
\centering{
\fbox{
\begin{minipage}[ht!]{150mm}
\footnotesize
\renewcommand{\baselinestretch}{1.5}
\resetline
\begin{tabbing}
aaaaA\=aaaaaA\=aaaaaaA\kill
{\bf Initial state of node $i$} \\
\line{L00} \> $\ms{period}_{i}$, initially set to a constant; \\
$r_i$, a random value chosen in $(0,1]$; 
$a_i$, the attribute value;   \\
$\ms{slice}_i \gets \bot$, the slice $i$ belongs to;
${\mathcal N_i}$, the view; \\
$\ms{gain}_{j'}$, a real value indicating the gain achieved by exchanging with
$j'$; \\
$\ms{gain-max} = 0$, a real.
\\ ~ \\

%{\bf Active thread at node $i$} \\
%\line{L02} \> $\act{wait}(\ms{period_i})$ \\
%~(\ref{L02}a) \> $\act{recompute-view}()_i$ \\
%\line{L03} \> {\bf for} $j' \in {\mathcal N}_i$ \\
%~(\ref{L03}a) \> \T {\bf if} $\ms{gain}_{j'} \geq \ms{gain-max}$ {\bf then} \\
%~(\ref{L03}b) \> \T \T $\ms{gain-max} \gets \ms{gain}_{j'}$ \\
%~(\ref{L03}c) \> \T \T $j \gets j'$ \\
%~(\ref{L03}d) \> {\bf end for} \\
%\line{L04} \> $\act{send}(\lit{REQ}, r_i, a_i)$ to $j$ \\
%~(\ref{L04}a) \> $\act{recv}(\lit{ACK}, r_j')$ from $j$ \\
%\line{L05} \> $r_j \gets r_j'$ \\
%~(\ref{L05}a) \> {\bf if} $(a_j - a_i)(r_j - r_i) < 0$ {\bf then} \\
%~(\ref{L05}b) \> \T $r_i \gets r_j$ \\
%~(\ref{L05}c) \> \T $\ms{slice}_i \gets {\cal S}_{l,u}$ such that $l < r_i \leq u$ \\
% ~ \\

{\bf Active thread at node $i$} \\
\line{M01} \> $\act{wait}(\ms{period_i})$ \\
\line{M02} \> $\act{recompute-view}()_i$ \\
\line{M03} \> {\bf for} $j' \in {\mathcal N}_i$ \\
\line{M04} \> \T {\bf if} $\ms{gain}_{j'} \geq \ms{gain-max}$ {\bf then} \\
\line{M05} \> \T \T $\ms{gain-max} \gets \ms{gain}_{j'}$ \\
\line{M06} \> \T \T $j \gets j'$ \\
\line{M07} \> {\bf end for} \\
\line{M08} \> $\act{send}(\lit{REQ}, r_i, a_i)$ to $j$ \\
\line{M09} \> $\act{recv}(\lit{ACK}, r_j')$ from $j$ \\
\line{M10} \> $r_j \gets r_j'$ \\
\line{M11} \> {\bf if} $(a_j - a_i)(r_j - r_i) < 0$ {\bf then} \\
\line{M12} \> \T $r_i \gets r_j$ \\
\line{M13} \> \T $\ms{slice}_i \gets {\cal S}_{l,u}$ such that $l < r_i \leq u$ \\
 ~ \\

%% with period, the other nodes must send more frequently
%~(\ref{L01})  \> $\act{wait}(\ms{period_i})$ \\
%~(\ref{L02})  \> ${\mathcal N}_i \gets \act{recompute-view}()_i$ \\
%~(\ref{L03})  \> {\bf for} $j' \in {\mathcal N}_i$ \\
%~(\ref{L04}') \> \T {\bf if} $\lit{dist}(a_{j'},b) < dmin$ {\bf then} \\
%~(\ref{L05}') \> \T \T $\ms{dist-min} \gets \lit{dist}(a_{j'},b)$ \\
%~(\ref{L06}')  \> \T \T $j \gets j'$ \\
%\line{L11}    \> \T {\bf if} $a_{j'} \leq a_i$ {\bf then} $\ell_i \gets \ell_i + 1$ \\
%\line{L12}    \> \T $g_i \gets g_i + 1$ \\
%~(\ref{L07})  \> $\act{send}(\lit{REQ},r_i)$ to $j$ \\
%~(\ref{L08})  \> $\act{recv}(\lit{ACK},r_j)$ from $j$ \\
%\line{L13}   \> {\bf if} $a_{j'} \leq a_i$ {\bf then} $\ell_i \gets \ell_i + 1$ \\
%\line{L14}   \> $g_i \gets g_i + 1$ \\
%\line{L15}   \> $r_i \gets \ell_i / g_i$ \\
%\line{L16}   \> $\ms{slice} \gets {\cal S}_{l,u}$ such that $l < r_i \leq u$\\
% ~ \\

{\bf Passive thread at node $i$ activated upon reception} \\
\line{R01} \> $\act{recv}(\lit{REQ}, r_j, a_j)$ from $j$ \\
\line{R02} \> $\act{send}(\lit{ACK}, r_i)$ to $j$ \\
\line{R03} \> {\bf if} $(a_j - a_i)(r_j - r_i) < 0$ {\bf then} \\
\line{R04} \> \T $r_i \gets r_j$ \\
\line{R05} \> \T $\ms{slice}_i \gets {\cal S}_{l,u}$ such that $l < r_i \leq u$
 
\end{tabbing}
\normalsize
\end{minipage} 
}
\caption{Dynamic ordering by exchange of random values.}
\label{alg:rand}
}
\end{figure}

The algorithm for exchanging random values from node $i$ starts by measuring 
the ordering that can be gained by swapping with each neighbor (Lines~\ref{M03}--\ref{M07}).
Then, $i$ chooses the neighbor $j \in {\mathcal N}_i$ that maximizes gain 
$G_{i,k}$ for any of its neighbor $k$. Formally, $i$ finds 
$j\in {\mathcal N}_i$ such that for any $k\in{\mathcal N}_i$, we have
\begin{eqnarray}
G_{i,j}(t+1) &\geq& G_{i,k}(t+1).  \label{eq:gainmax}
\end{eqnarray}
Using the definition of $G_{i,j}$ in Equation (\ref{eq:gain}), Equation (\ref{eq:gainmax}) is 
equivalent to 
\begin{eqnarray}
\ell\alpha_i(t) \ell\rho_j(t) + \ell\alpha_j(t) \ell\rho_i(t) - \ell\alpha_j(t)
\ell\rho_j(t)  &\geq& \ell\alpha_i(t) \ell\rho_k(t) + \ell\alpha_k(t) \ell\rho_i(t)
- \ell\alpha_k(t) \ell\rho_k(t). \notag
\end{eqnarray}
\noindent In Figure~\ref{alg:rand} of node $i$, we refer to $\ms{gain}_j$ as the value of
$\ell\alpha_i(t) \ell\rho_j(t) + \ell\alpha_j(t) \ell\rho_i(t) - \ell\alpha_j(t)
\ell\rho_j(t)$.
%\begin{eqnarray}
%G_{i,j}(t+1) &\geq& G_{i,k}(t+1), \notag \\
%\ell\alpha_i \ell\rho_j(t) + \ell\alpha_j \ell\rho_i(t) - \ell\alpha_j \ell\rho_j(t)  &\geq& \ell\alpha_i \ell\rho_k(t) + \ell\alpha_k \ell\rho_i(t) - \ell\alpha_k \ell\rho_k(t). \Box\notag
%\end{eqnarray}

%%this $\ms{gain}$ (i.e., 
%neighbor $j$ whose swapping minimizes the local disorder measure of $i$).
From this point on, $i$ exchanges its random value $r_i$ with the random value $r_j$ of 
node $j$ (Line~\ref{M10}). The passive threads are executed upon reception of a message.
In Figure~\ref{alg:rand}, when $j$ receives the random value $r_i$ of node 
$i$, it sends back its own random value $r_j$ for the exchange to occur (Lines~\ref{R01}--\ref{R02}).
Observe that the attribute value of $i$ is also sent to $j$, so that $j$ can
check if it is correct to exchange before updating its own random number (Lines~\ref{R03}--\ref{R04}). Node
$i$ does not need to receive attribute value $a_{j}$ of $j$, since $i$ already has this 
information in its view and the attribute value of a node never changes over time.

\begin{figure}
\centering{
\fbox{
\begin{minipage}[b]{150mm}
\footnotesize
\renewcommand{\baselinestretch}{1.5}
\resetline
\begin{tabbing}
aaaaA\=aaaaaA\=aaaaaaA\kill
%{\bf Initial state of node $i$} \\
%\line{L00} \> $r_i$, a random value chosen in $(0,1]$. 
%$a_i$, the attribute\\ value. 
%${\mathcal N_i}$, the view.
%\\ ~ \\
%
{\bf Active thread at node $i$} \\
\line{V01} \> {\bf for} $j' \in {\mathcal N}_{i}$ {\bf do} $t_{j'} \gets t_{j'} + 1$ {\bf end for} \\
\line{V02} \> $j \gets j'': t_{j''} = \lit{max}_{j'\in {\mathcal N}_i}(t_{j'})$ \\
\line{V03} \> $\act{send}(\lit{REQ'}, {\mathcal N}_i \setminus \{e_j\} \cup \{\tup{i, 0, a_i, r_i}\})$ to $j$ \\
\line{V04} \> $\act{recv}(\lit{ACK'}, {\mathcal N}_{j})$ from $j$ \\
\line{V05} \> $\ms{duplicated-entries} = \{e: e.id \in {\mathcal N}_j \cap {\mathcal N}_i\}$ \\
\line{V06} \> ${\mathcal N}_i \gets {\mathcal N}_i \cup ({\mathcal N}_j \setminus \ms{duplicated-entries} \setminus \{e_i\})$ \\
~ \\

{\bf Passive thread at node $i$ activated upon reception} \\
\line{RV01} \> $\act{recv}(\lit{REQ'}, {\mathcal N}_{j})$ from $j$ \\
\line{RV03} \> $\act{send}(\lit{ACK'}, {\mathcal N}_{i})$ to $j$ \\
\line{RV04} \> $\ms{duplicated-entries} = \{e \in {\mathcal N}_j: e.id \in {\mathcal N}_j \cap {\mathcal N}_i\}$ \\
\line{RV05} \> ${\mathcal N}_i \gets {\mathcal N}_i \cup ({\mathcal N}_j \setminus \ms{duplicated-entries})$ \\
%{\bf last line says that we updates views and keep timestamp of i'e entries, even if entries are duplicated in Ni and Nj}
%\line{RV05} \> ${\mathcal N}_i \gets {\mathcal N}_i \cup ({\mathcal N}_j \setminus ({\mathcal N}_j \cap {\mathcal N}_i))$

\end{tabbing}
\normalsize
\end{minipage} 
}
\caption{$\lit{recompute-view()}$: procedure used to update the view based on a simple variant of the Cyclon algorithm.}
\label{alg:rcyclon}
}
\end{figure}

\subsection{Analysis of Slice Misplacement}
\label{sec:anal}

In mod-JK, as in JK, the current random number
$r_{i}$ of a node $i$ determines the slice $s_{i}$ of the node. The
objective of both algorithms is to reduce the global disorder as quickly as
possible.  Algorithm mod-JK consists in choosing one neighbor among the possible neighbors
that would have been chosen in JK, plus the GDM of JK has been shown to fit an exponential 
decrease~\cite{JK06}. Consequently mod-JK experiences also an exponential decrease of the global disorder.
Eventually, JK and mod-JK 
ensure that the disorder has fully disappeared.
However, the accuracy of the slices heavily depends on the uniformity of the random value
spread between 0 and 1.  It may happen, that the distribution of the random values
is such that some peers decide upon a wrong slice. Even more problematic is the fact that this situation
is unrecoverable unless a new random value is drawn for all nodes. 
This may be considered as an inherent limitation of
the approach.
For example, consider a system of size 2, where nodes 1 and 2 have the
random values $r_1=0.1$, $r_2=0.4$. If we are interested in creating
two slices of equal size, the first slice will be of size 2 and the second
of size zero, even after perfect ordering of the random values.

Therefore, an important step is to characterize the inaccuracy of the 
uniform distribution  to access the potential impact on
the  slice assignment
resulting from the fact that uniformly generated random numbers are
not distributed perfectly evenly throughout the domain.
First of all, consider a slice $S_p$ of length $p$.
In a network of $n$ nodes,
the number of nodes that will fall into this slice is a random variable $X$
with a binomial distribution with parameters $n$ and $p$.
The standard deviation of $X$ is therefore $\sqrt{np(1-p)}$.
This means that the relative proportional expected difference from the
mean (i.e., $np$) can be approximated as $\sqrt{(1-p)/(np)}$, which is very large
if $p$ is small, in fact, goes to infinity as $p$ tends to zero,
although a very large $n$ compensates for this effect.
For a ``normal'' value of $p$, and a reasonably large network, the variance
is very low however.

To stay with this random variable,
the following result bounds, with high probability, its deviation from its
mean.
%Consequently, this bounds (w.h.p.) the number of nodes that can be misplaced in the system.

\begin{lemma}
For any $\beta \in (0,1]$, a slice $S_p$ of length  $p \in (0,1]$ has a
number of peers 
$X \in [(1-\beta)np, (1+\beta)np]$ with probability at least 
$1-\epsilon$ as long as $p  \geq \frac{3}{\beta^2 n} \ln (2/\epsilon)$.
\end{lemma}
\begin{proof}
The way nodes choose their random number is like drawing $n$ times, with replacement and
independently uniformly at random, a value in the interval $(0,1]$. 
Let $X_1, ..., X_n$ be the $n$ corresponding independent identically distributed random 
variables such that:
\begin{equation}
\left\{\begin{array}{ll}
X_i &= 1 \text{~if the value drawn by node $i$ belongs to~}S_p\text{~and~}\notag
\\
X_i &= 0 \text{~otherwise.}
\end{array}\right.
\end{equation}

We denote $X= \sum_{i=1}^n X_i$ the number of elements of interval $S_p$ drawn 
among the $n$ drawings. 
The expectation of $X$ is  $np$. 
From now on we compute the probability that a bounded portion of the expected 
elements are misplaced.
Two Chernoff bounds~\cite{MR95} give:
\begin{equation}
\left.\begin{array}{cc}
\Pr[X\geq (1+\beta)np] &\leq e^{-\frac{\beta^2 np}{3}} \notag \\
\Pr[X\leq (1-\beta)np] &\leq e^{-\frac{\beta^2 np}{2}} \notag
\end{array}\right\} \\
\Rightarrow \Pr[|X-np| \geq \beta np] \leq 2e^{-\frac{\beta^2 np}{3}},
\end{equation}

\noindent with $0 < \beta \leq 1$. 
That is, the probability that more than ($\beta$ time the number expected) elements are 
misplaced regarding to interval $S_p$ is bounded by $2e^{-\frac{\beta^2 n
p}{3}}$. We want this to be at most $\epsilon$. This yields the result.
\end{proof}
%
%The previous result shows that the number of nodes allocated to a given slice
%is not far from the amount is should be.
%
%%MJ Note that you draw the wrong conclusion (commented out), as the simple fact I included
% shows that if p is small then we are in trouble, and also if n is
% not that large. The Chernoff shows the same, but in a more obscure way.
% I repeat that these bounds are useful only if one wants to prove some
% complexity results, and these bounds are not that tight anyway to be
% very useful directly.
% Besides, the "theorem" is a simple application of the Chernoff bound, and
% is rather basic. I renamed it to "lemma", and I also feel the proof can
% be significantly shortened. To something like: "simple application of
% the Chernoff bound".
% I left the derivation in though, with these notes...

To measure the effect discussed above during the simulation experiments, 
we introduce the slice 
disorder measure (SDM) as the sum over all nodes $i$ of the
distance between the slice $i$ actually belongs to and the slice $i$ believes
it belongs to.
For example (in the case where all slices have the same size), 
if node $i$ belongs to the $1^{st}$ slice (according to its attribute 
value) while it thinks it belongs to the $3^{rd}$ slice (according to its rank 
estimate) then the distance for node $i$ is $|1-3| = 2$.
Formally, for any node $i$, let $S_{u_i, l_i}$ be the actual correct slice 
of node $i$ and let $S_{\hat{u_i}, \hat{l_i}}(t)$ be the slice $i$ estimates
as its slice at time $t$.  The slice disorder measure is defined as:
$$\ms{SDM}(t) = \sum_{i}\frac{1}{u_i - l_i}\left|\frac{u_i+l_i}{2}-\frac{\hat{u_i}+\hat{l_i}}{2}\right|.$$
$\ms{SDM}(t)$ is minimal (equals 0) if for all nodes $i$, 
we have $S_{\hat{u_i}, \hat{l_i}}(t) = S_{u_i, l_i}$.
%
%
%This
%interval
%is divided in $k$ subintervals $I_1, ..., I_k$ of length $\Delta_1, ..., \Delta_k \in (0,1]$, respectively.  
%Let $\Delta$ be the minimal length of these intervals, formally $\Delta = min_{\forall j} \{p\}$.
%Without loss of generality fix $j$ such that $p$ is one of these subintervals.
%
%
%
%
%We then bound (using additive bounds), with high probability, the number of misplaced 
%elements with respect to all subintervals.
%\begin{eqnarray}
%P &=& \sum_{j=1}^k \Pr[|X-np| > \beta np] \leq 2k e^{-\frac{\beta^2n\Delta}{3}}. \notag
%\end{eqnarray}
%
%We want $P \leq \epsilon$:
%\begin{eqnarray}
%& 2k e^{-\frac{\beta^2 n \Delta}{3}} \leq \epsilon, \notag \\
%\Rightarrow & -\frac{\beta^2n\Delta}{3} \leq \ln{\frac{\epsilon}{2k}}, \notag \\
%\Rightarrow & \Delta \geq \frac{3}{\beta^2 n}\ln{\frac{2k}{\epsilon}}.\Box \notag
%\end{eqnarray}
%\end{proof}
%
%\begin{corollary}
%For any $\beta \in (0,1]$, any slice of length $\Delta_i$ has a number of peers $n_i \in [(1-\beta)n\Delta_i, (1 +\beta)n\Delta_i]$ with high probability (of at least $1-1/n$) as long as  $\ms{min}_j p \geq \frac{3}{\beta^2 n} \ln{(2nk)}$.
%\end{corollary}

In fact, it is simple to show that, in general, the probability of  dividing $n$ peers into two slices of the same size is less than $\sqrt{2/n\pi}$. This value is very small even for moderate values of  $n$. Hence, it is highly possible that the random number distribution  does not lead to a perfect division into slices. 

\subsection{Simulation Results}\label{sec:simu2}

We present simulation results using PeerSim~\cite{JMB04}, using a simplified
cycle-based simulation model, where all messages exchanges are atomic,
so messages never overlap.
First, we compare the performance of the two algorithms: JK and mod-JK.
Second, we study the impact of concurrency that is ignored by the cycle-based
simulations.

\subsubsection{Performance Comparison}
%To compare the performance of our algorithm, 
%mod-JK, with JK,
%we implemented both approaches in a cycle-based 
%simulator using PeerSim~\cite{JMB04}.
We compare the time taken by these algorithms 
to sort the random values according to the attribute values (i.e., the node with 
the $j^{th}$ largest attribute value of the system value obtains the $j^{th}$ random value).
In order to evaluate the convergence speed of each algorithm, we use the slice
disorder measure as defined in Section~\ref{sec:anal}.

\begin{figure*}
  \begin{center}
    \subfigure[]
    { 
      \label{fig:convergence2}
      \resizebox{2.6in}{!}{\includegraphics[scale=0.5,angle=270,clip=true]{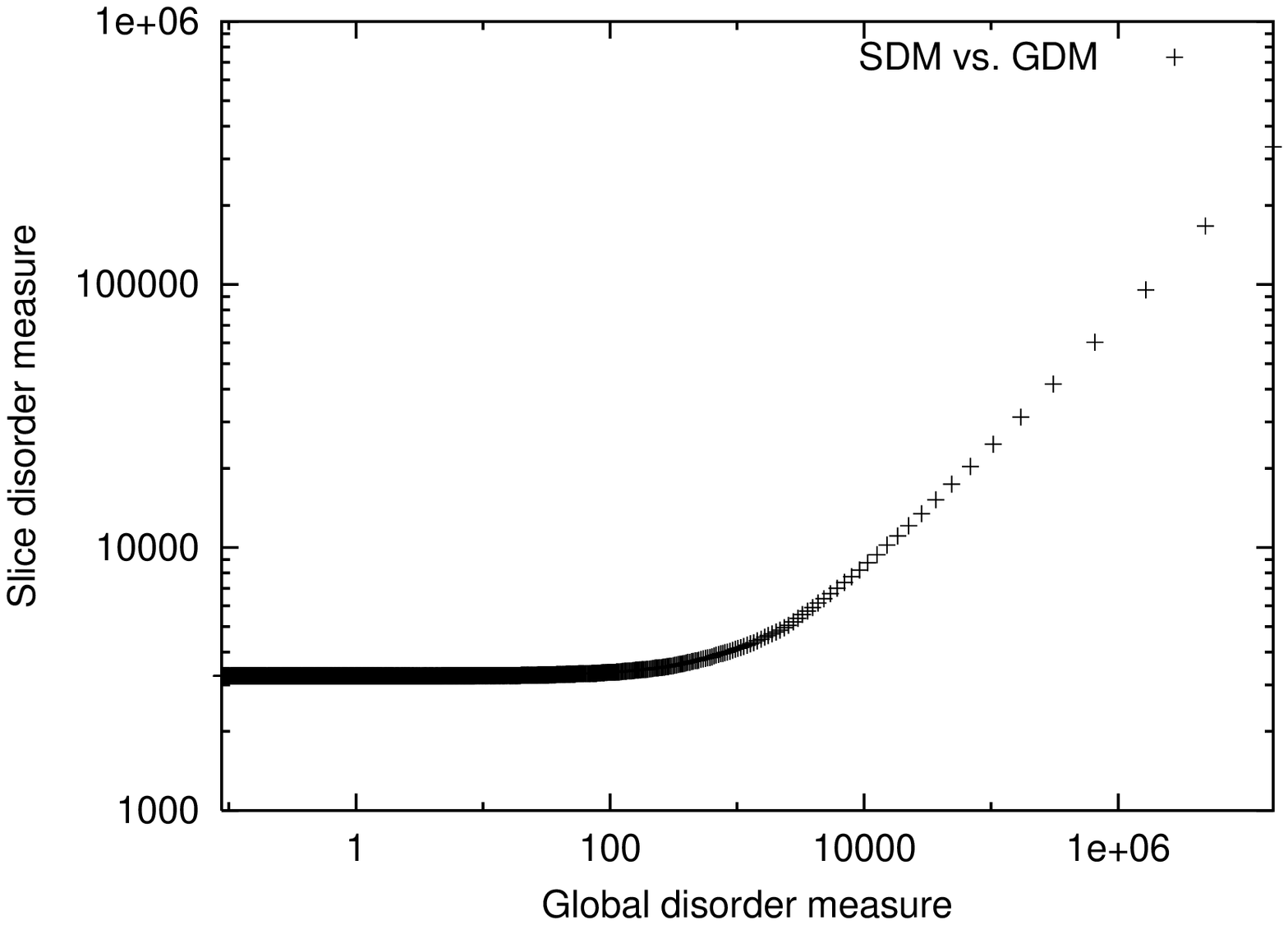}}
    }
    \subfigure[]
    {
    \label{fig:convergence1}
          \resizebox{2.6in}{!}{\includegraphics[scale=0.5,angle=270,clip=true]{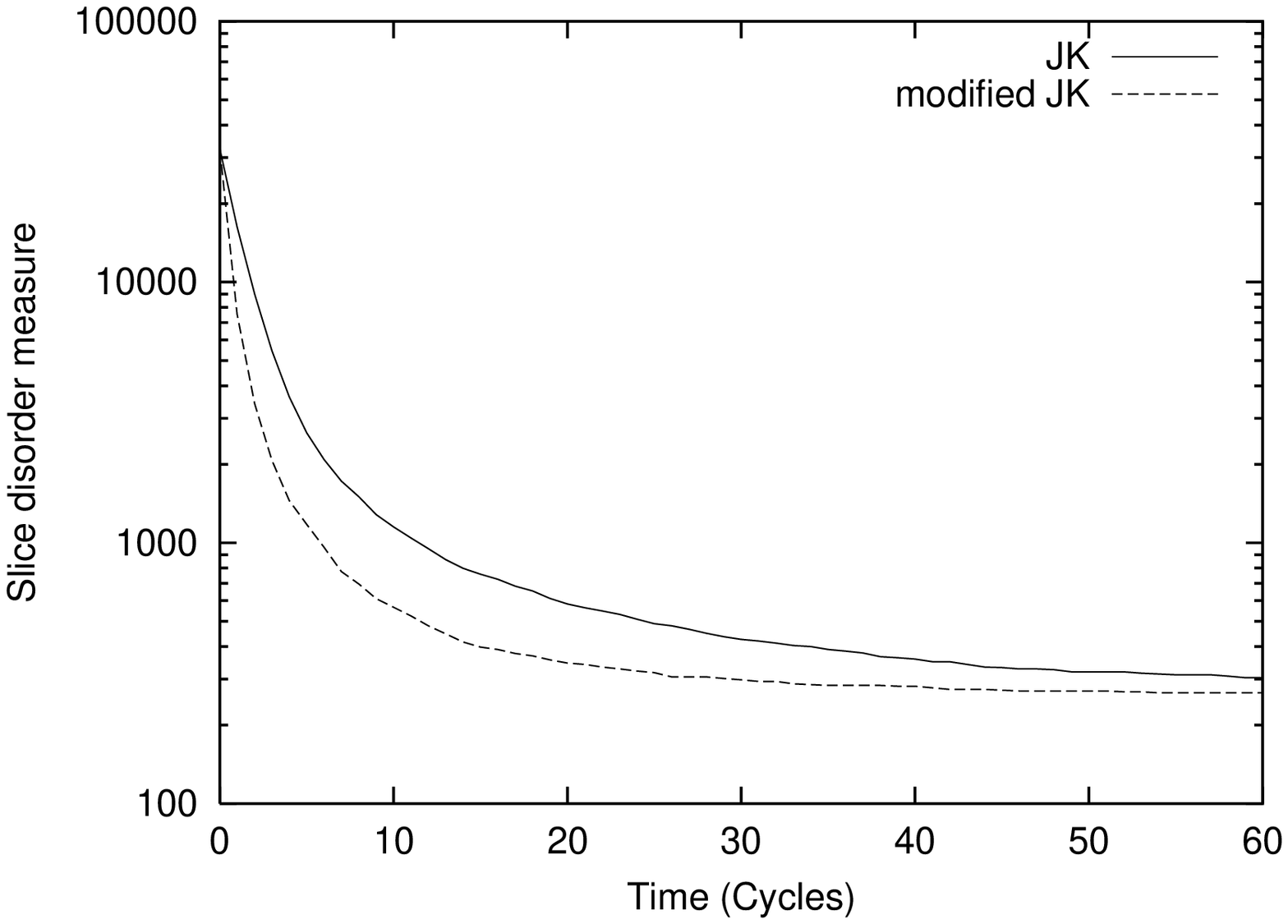}}
    }
    \subfigure[]
    { 
      \label{fig:swap}
      \resizebox{2.6in}{!}{\includegraphics[scale=0.5,angle=0,clip=true,bb=40 10 260 200,viewport=0 0 260 210
      ]{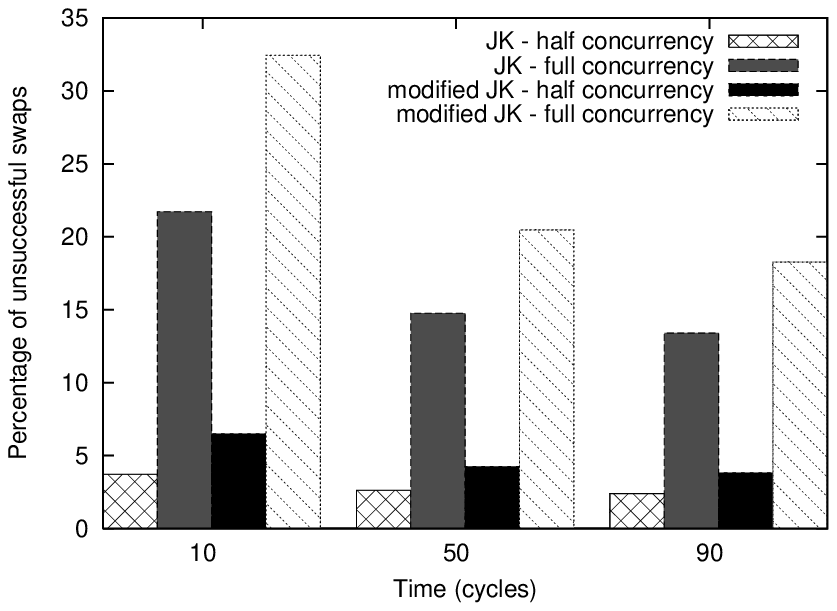}}
    } 
    \subfigure[]
    { 
      \label{fig:sdm_count}
      \resizebox{2.6in}{!}{\includegraphics[scale=0.5,angle=270,origin=c,clip=true]{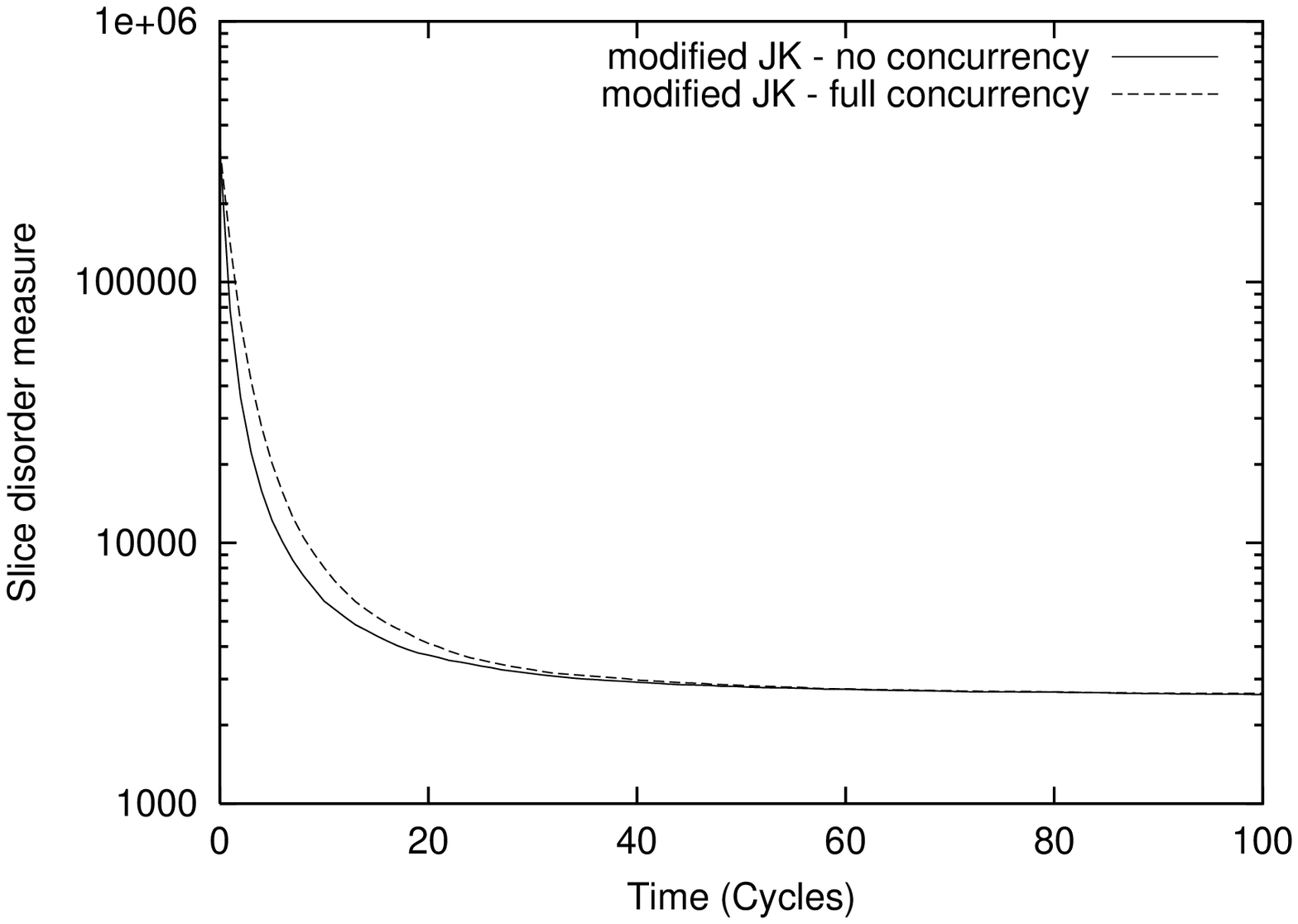}}
    }    
    \caption{
       (a) Evolution of the global disorder measure over time.
       (b) Slice disorder measure over time.
       (c) Percentage of unsuccessful swaps in the ordering algorithms.
       (d) Convergence speed under high concurrency.}
  \end{center}
\end{figure*}

We simulated $10^4$ participants in 100 equally sized slices (when unspecified), each with 
a view size $c=20$.  Figure~\ref{fig:convergence2} illustrates the difference between the global disorder
measure and the slice disorder measure while Figure~\ref{fig:convergence1} presents the evolution of
the slice disorder measure over time for JK, and mod-JK.

Figure~\ref{fig:convergence2} shows the different speed at which 
the global disorder measure and the slice disorder measure converge.
When values are sufficiently large, the GDM and SDM seem tightly related: if
GDM increases then SDM increases too.  Conversely, there is a significant difference between 
the GDM and SDM when the values are relatively low: the GDM reaches 0 while the SDM is lower bounded by 
a positive value.  This is because the algorithm does lead to a totally
ordered set of nodes, while it still does not associate each node with its
correct slice. Consequently the GDM is not sufficient to rightly estimate the
performance of our algorithms.

Figure~\ref{fig:convergence1} shows the slice disorder measure to compare
the convergence speed of our algorithm to that of JK with 10 equally sized slices.
Our algorithm converges significantly faster than JK.
Note that none of the algorithm reaches zero SDM, since they are both based on
the same idea of sorting randomly generated values.
Besides, since they both used an identical set of randomly generated values,
both converge to the same SDM.

\subsubsection{Concurrency}

The simulations are cycle-based and at each cycle an 
algorithm step is done atomically so that no other execution is concurrent.
%a node sends and receives messages in an atomic 
%manner---while no other messages are in transit.  
More precisely, the algorithms are simulated such that in each cycle, each node updates its view 
before sending its random value or its attribute value.  Given this implementation, the 
cycle-based simulator does not allow us to realistically simulate concurrency, and a
drawback is that view is up-to-date when a message is sent.  In the following 
we artificially introduce concurrency (so that view might be out-of-date) into the 
simulator and show that it has only a slight impact on the convergence speed.

%This parameter is not realistic in case the
%goal is to converge as fast as possible, especially if messages
%are continuously sent for this purpose.  Let $\delta$ be the time a message takes to reach 
%its destination.  If a large portion of nodes send a message in a 
%period of $\delta$ time units then it is likely that at least two overlapping messages target the 
%same node.
%Here, we simulate an artificial concurrency to see the impact on convergence speed.

%This implementation choice presents two side-effects in case of high concurrency injection, 
%proper to ordering algorithms.  
%First, there might be a large number of useless messages.  Second, the convergence
%can slow down compare to the non-concurrent case.  
%
Introducing concurrency might result in some problems because of
%These two problems arise from 
the potential staleness of views: unsuccessful swaps due to useless messages.  
Technically, the view of node $i$ might 
indicate that $j$ has a random value $r$ while this value is no longer 
up-to-date.  This happens if $i$ has lastly updated its view before $j$ swapped
its random value with another $j'$.
Moreover, due to asynchrony, it could happen that by the time a message is received
this message has become useless.  Assume that node $i$ sends its random value $r_i$ to $j$
in order to obtain $r_j$ at time $t$ and $j$ receives it by time $t+\delta$. 
With no loss of generality assume $r_i > r_j$.  Then if $j$ swaps its random value with
$j'$ such that $r_j'>r_i$ between time $t$ and $t+\delta$, then the message of $i$ 
becomes \emph{useless} and the expected swap does not occur (we call this an \emph{unsuccessful swap}).
%
%
%When $i$ sends a message 
%$m_i$ to a targeted neighbor $j$, then between the time $i$ updates its view and the time 
%$j$ receives the message $m_i$, $j$ may swap its own random value with a distinct node $j'$ (upon 
%reception of message $m_{j'}$ from $j'$). 
%%
%In case the message from $j'$ results in ordering $j$ regarding to $i$, then $j$ will
%ignore the message from $i$.  This scenario occurs if initially $r_{j'} \leq r_i \leq r_j$
%and $a_j < a_i$ or $r_j \leq r_i \leq r_{j'}$ and $a_i < a_j$. 
%Consequently, the random value of $j$ that $i$ knows of is stale, and message $m_i$ does not
%result in any swap, $j$ ignores it.

%
%\begin{figure*}
%  \begin{center}
%    \caption{
%    }
%  \end{center}
%\end{figure*}

Figure~\ref{fig:sdm_count} indicates the impact of concurrent message exchange
on the convergence speed while
Figure~\ref{fig:swap} shows the amount of useless messages that are sent.
%shows the percentage of messages that are ignored during
%simulations of the ordering algorithms (JK and mod-JK) over the total number 
%of messages.
Now, we explain how the concurrency is simulated.
Let the \emph{overlapping messages} be a set of messages that mutually overlap: 
it exists, for any couple of overlapping messages, at least one instant at 
which they are both in-transit.  
For each algorithm we simulated \textit{(i)}~full concurrency: in a given 
cycle, all messages are overlapping messages; and \textit{(ii)}~half 
concurrency: in a given cycle, each message is an overlapping message with 
probability $\frac{1}{2}$.
Generally, we see that increasing the concurrency increases the number of
useless messages. Moreover, in the modified version of JK, more messages are 
ignored than in the original JK algorithm.  This
is due to the fact that some nodes (the most misplaced ones) are more likely 
targeted which increases the number of concurrent messages arriving at the 
same nodes.  Since a node $i$ ignored more likely a message when it receives
more messages during the same cycle, it comes out that concentrating
message sending at some targets increases the number of useless messages.

Figure~\ref{fig:sdm_count} compares the convergence speed under full concurrency 
and no concurrency.  We omit the curve of half-concurrency since it would have been 
similar to the two other curves.  Full-concurrency 
impacts on the convergence speed very slightly.
%While in the non-concurrent case all messages result
%in an ordering gain, in the concurrent case, some of them may not.
%
%These side-effects are absent in the ranking algorithm (Section~\ref{sec:ranking}) because the 
%information conveyed in message from each node $i$ is altered 
%independently from any message receipt.
%Indeed, either attribute values slightly change over time (if it 
%represents battery, or available storage space), or do not change 
%at all, while an ordering algorithm makes nodes exchange random values
%that can be drastically modified (swapped) upon reception of another message.

\section{Dynamic Ranking by Sampling of Attribute Values}\label{sec:ranking}
\label{dynamicranking}
In this section we propose an alternative algorithm for the distributed slicing
problem. This algorithm circumvents some of the problems identified in
the previous approach by continuously ranking nodes based on observing attribute value
information. 
Random values no longer play a role, so non-perfect uniformity in the random value
distribution is no longer a problem.
Besides, this algorithm
is not sensitive to churn even if it is correlated with attribute values. 

In the remaining part of the report we refer to this new algorithm 
as the ranking algorithm while referring to JK and mod-JK as the ordering algorithms.
Here, we elaborate on the drawbacks arising from the ordering algorithms 
relying on the use of random values that are solved by the ranking approach.

\paragraph{Impact of attribute correlated with dynamics.}
As already mentioned, the ordering algorithms
rely on the fact that random values are uniformly distributed.
However, if the attribute values are not constant but correlated with
the dynamic behavior of the system, the distribution of random values may change
from uniform to  skewed quickly.
For instance, assume that each node maintains an attribute
value that represents its own lifetime. 
Although the algorithm is able to quickly sort random values, so
nodes with small lifetime will obtain the small random values, it
is more likely that these nodes leave the system sooner than other nodes.
This results in
a higher concentration of high random values and a large population of the nodes
wrongly estimate themselves as being part of the higher slices.
%Similar situations arise when
%attribute values represent other dynamic features such as
%the remaining storage space of a node, etc.

\paragraph{Inaccurate slice assignments.}

As discussed in previous sections in detail, slice assignments will
typically be imperfect even when the random values are perfectly
ordered.
Since the ranking approach does not rely on ordering random nodes,
this problem is not raised: the algorithm guarantees eventually
perfect assignment in a static environment.

\paragraph{Concurrency side-effect.}
In the previous ordering algorithms, a non negligible amount of messages are sent unnecessarily.
The concurrency of messages has a drastic effect on the number of useless messages
as shown previously, slowing down convergence.
In the ranking algorithm
concurrency has no impact on convergence speed because all 
received messages are taken in account. 
This is because the information encapsulated in a message (the attribute value of a node)
is guaranteed to be up to date, as long as  the attribute values are constant,
or at least change slowly.

\subsection{Ranking Algorithm Specification}

The pseudocode of the ranking algorithm is presented in Figure~\ref{alg:attr}.
As opposed to the ordering algorithm of the previous section, the ranking
algorithm does not assign random initial unalterable values as candidate ranks.
Instead, the ranking algorithm improves its rank estimate each
time a new message is received.

The ranking algorithm works as follows.
Periodically each node $i$ updates its view ${\mathcal N}_i$ following 
an underlying protocol that provides a uniform random sample (Line~\ref{N02}); later, we simulate the algorithm using
the variant of Cyclon protocol presented in Section~\ref{sec:modifiedjk}.
Node $i$ computes its rank estimate (and hence its slice)
by comparing the attribute value of its neighbors to its own attribute value.
This estimate is set to the ratio of the number of nodes with a lower 
attribute value that $i$ has seen over the total number of nodes $i$ has seen (Line~\ref{N14}).
Node $i$ looks at the normalized rank estimate of all its neighbors.
Then, $i$
selects the node $j_{1}$ closest to a slice boundary (according to the rank 
estimates of its neighbors).
Node $i$ selects also a random neighbor $j_{2}$ among its view (Line~\ref{N11}).
When those two nodes are selected, $i$ sends an update message, denoted by a flag $\lit{UPD}$, 
to $j_{1}$ and $j_{2}$
containing its attribute value (Line~\ref{N12}--\ref{N13}).

The reason why a node close to the slice boundary is selected as one of the
contacts is that such nodes need more samples to accurately determine
which slice they belong to (subsection~\ref{sec:analysis2} shows this point).
This technique introduces a bias towards them, so they receive more messages.

Upon reception of a message from node $i$, the passive threads of $j_{1}$ and 
$j_{2}$ are activated so that $j_{1}$ and $j_{2}$ compute their new rank estimate 
$r_{j_{1}}$ and $r_{j_{2}}$.  The estimate of the slice a node belongs to,
follows the computation of the rank estimate.
Messages are not replied, communication is one-way, resulting in identical message
complexity to JK and mod-JK.

\begin{figure}
\centering{
\fbox{
\begin{minipage}[t]{150mm}
\footnotesize
\renewcommand{\baselinestretch}{1.5}
\resetline
\begin{tabbing}
aaaaA\=aaaaaA\=aaaaaaA\kill
{\bf Initial state of node $i$} \\
\line{N00} \> $\ms{period}_{i}$, initially set to a constant;
$r_i$, a value in $(0,1]$; \\
$a_i$, the attribute value;
$b$, the closest slice boundary to node $i$; \\
$g_i$, the counter of encountered attribute values;
$l_i$, the counter \\
of lower attribute values;
$\ms{slice}_i \gets \bot$;
${\mathcal N_i}$, the view.
\\ ~ \\
 
{\bf Active thread at node $i$} \\

%% with period, the other nodes must send more frequently
%~(\ref{L02}b)  \> $\act{wait}(\ms{period_i})$ \\
%~(\ref{L02}c)  \> $\act{recompute-view}()_i$ \\
%~(\ref{L02}d)  \> $\ms{dist-min} \gets \infty$ \\
%~(\ref{L03}e)  \> {\bf for} $j' \in {\mathcal N}_i$ \\
%~(\ref{L03}f)    \> \T $g_i \gets g_i + 1$ \\
%~(\ref{L03}g)   \> \T {\bf if} $a_{j'} \leq a_i$ {\bf then} $\ell_i \gets \ell_i + 1$ \\
%~(\ref{L03}h) \> \T {\bf if} $\lit{dist}(a_{j'},b) < \ms{dist-min}$ {\bf then} \\
%~(\ref{L03}i) \> \T \T $\ms{dist-min} \gets \lit{dist}(a_{j'},b)$ \\
%~(\ref{L03}j) \> \T \T $j_{1} \gets j'$ \\
%~(\ref{L03}k) \> {\bf end for} \\
%~(\ref{L03}l) \> Let $j_{2}$ be a random node of ${\mathcal N}_i$ \\
%~(\ref{L04}b)  \> $\act{send}(\lit{UPD},r_i)$ to $j_{1}$ \\
%~(\ref{L04}c)  \> $\act{send}(\lit{UPD},r_i)$ to $j_{2}$ \\
%%~(\ref{L08})  \> $\act{recv}(\lit{ACK},r_j)$ from $j$ \\
%%\line{L13}   \> {\bf if} $a_{j'} \leq a_i$ {\bf then} $\ell_i \gets \ell_i + 1$ \\
%%\line{L14}   \> $g_i \gets g_i + 1$ \\
%~(\ref{L05}d)   \> $r_i \gets \ell_i / g_i$ \\
%~(\ref{L05}e)   \> $\ms{slice} \gets {\cal S}_{l,u}$ such that $l < r_i \leq u$\\
% ~ \\

% with period, the other nodes must send more frequently
\line{N01}  \> $\act{wait}(\ms{period_i})$ \\
\line{N02}  \> $\act{recompute-view}()_i$ \\
\line{N03}  \> $\ms{dist-min} \gets \infty$ \\
\line{N04}  \> {\bf for} $j' \in {\mathcal N}_i$ \\
\line{N05}    \> \T $g_i \gets g_i + 1$ \\
\line{N06}   \> \T {\bf if} $a_{j'} \leq a_i$ {\bf then} $\ell_i \gets \ell_i + 1$ \\
\line{N07} \> \T {\bf if} $\lit{dist}(a_{j'},b) < \ms{dist-min}$ {\bf then} \\
\line{N08} \> \T \T $\ms{dist-min} \gets \lit{dist}(a_{j'},b)$ \\
\line{N09} \> \T \T $j_{1} \gets j'$ \\
\line{N10} \> {\bf end for} \\
\line{N11} \> Let $j_{2}$ be a random node of ${\mathcal N}_i$ \\
\line{N12}  \> $\act{send}(\lit{UPD},a_i)$ to $j_{1}$ \\
\line{N13}  \> $\act{send}(\lit{UPD},a_i)$ to $j_{2}$ \\
%~(\ref{L08})  \> $\act{recv}(\lit{ACK},r_j)$ from $j$ \\
%\line{L13}   \> {\bf if} $a_{j'} \leq a_i$ {\bf then} $\ell_i \gets \ell_i + 1$ \\
%\line{L14}   \> $g_i \gets g_i + 1$ \\
\line{N14}   \> $r_i \gets \ell_i / g_i$ \\
\line{N15}   \> $\ms{slice} \gets {\cal S}_{l,u}$ such that $l < r_i \leq u$\\
 ~ \\

%{\bf Passive thread at node $i$ activated upon reception} \\
%~(\ref{R01}) \> $\act{recv}(\lit{UPD},a_j)$ from $j$ \\
%%~(\ref{R02}) \> $\act{send}(\lit{ACK},r_i)$ to $j$ \\
%\line{R06}   \> {\bf if} $a_{j} \leq a_i$ {\bf then} $\ell_i \gets \ell_i + 1$ \\
%\line{R07}   \> $g_i \gets g_i + 1$ \\
%\line{R08}   \> $r_i \gets \ell_i / g_i$ \\
%~(\ref{R04}') \> $\ms{slice} \gets {\cal S}_{l,u}$ such that $l < r_i \leq u$

{\bf Passive thread at node $i$ activated upon reception} \\
\line{O01} \> $\act{recv}(\lit{UPD},a_j)$ from $j$ \\
%~(\ref{R02}) \> $\act{send}(\lit{ACK},r_i)$ to $j$ \\
\line{O02}   \> {\bf if} $a_{j} \leq a_i$ {\bf then} $\ell_i \gets \ell_i + 1$ \\
\line{O03}   \> $g_i \gets g_i + 1$ \\
\line{O04}   \> $r_i \gets \ell_i / g_i$ \\
\line{O05} \> $\ms{slice} \gets {\cal S}_{l,u}$ such that $l < r_i \leq u$

\end{tabbing}
\normalsize
\end{minipage} 
}
\caption{Dynamic ranking by exchange of attribute values.}
\label{alg:attr}
}
\end{figure}

\subsection{Theoretical Analysis}\label{sec:analysis2}

The following Theorem shows a lower bound on the probability 
for a node $i$ to accurately estimate the slice it belongs to.
This probability depends not only on the number of attribute exchanges
but also on the rank estimate of $i$.

\begin{theorem}
%Assume that node $i$ with normalized rank $p$, estimated in $\hat{p}$, executes 
%$\left( Z_\frac{\alpha}{2} \frac{ \sqrt{\hat{p}(1-\hat{p})} }{ d } \right)^2$
%steps of the algorithm, where $d$ is the distance between the rank estimate of $i$ and 
%the closest slice boundary, and $Z_\frac{\alpha}{2}$ represents the endpoints of the 
%confidence interval.  
%The slice estimate of $i$ is exact with confidence coefficient of $100(1 - \alpha)\%$.
%
Let $p$ be the normalized rank of $i$ and let $\hat{p}$ be its estimate.
For node $i$ to exactly estimate its slice with confidence coefficient of 
$100(1 - \alpha)\%$, the number of messages $i$ must receive is:
$$\left( Z_\frac{\alpha}{2} \frac{ \sqrt{\hat{p}(1-\hat{p})} }{ d } \right)^2,$$
\noindent
where $d$ is the distance between the rank estimate of $i$ and 
the closest slice boundary, and $Z_\frac{\alpha}{2}$ represents the endpoints of the 
confidence interval.  
\end{theorem}
\begin{proof}
%At each step of the algorithm execution, node $i$ draws uniformly at random 
%a sample in the network.
Each time a node receives a message, it checks whether or not the attribute value
is larger or lower than its own.  
Let $X_1, ..., X_k$ be $k$~$(k>0)$ independent identically distributed random variables described as follows. $X_j = 1$ 
with probability $\frac{i}{n} = p$ (indicating that the attribute value is lower) and 
$j \in \{1, ..., k\}$, otherwise $X_j = 0$ (indicating the attribute value is larger).
By the central limit theorem, we assume $k > 30$ and we approximate the distribution of $X = \sum_{j=1}^k X_j$
as the normal distribution. We estimate $X$ by $\hat{X} = \sum_{j=1}^k \hat{X}_j$
and $p$ by $\hat{p} = \frac{\hat{X}}{k}$.

We want a confidence coefficient with value $1-\alpha$.
Let $\Phi$ be the standard normal distribution function, and let 
$Z_{\frac{\alpha}{2}}$ be $\Phi^{-1}(1-\frac{\alpha}{2})$.
Now, by the Wald large-sample normal test in the binomial case, 
where the standard deviation of $\hat{p}$ is 
$\sigma(\hat{p}) = \frac{\sqrt{\hat{p}(1-\hat{p})}}{\sqrt{k}}$,
we have:
\begin{eqnarray}
\left|\frac{\hat{p}-p}{\sigma(\hat{p})}\right| &\leq Z_{\frac{\alpha}{2}} & \notag \\
\hat{p} - Z_{\frac{\alpha}{2}} \sigma(\hat{p}) &\leq p & \leq \hat{p} + Z_{\frac{\alpha}{2}}\sigma(\hat{p}). \notag
\end{eqnarray}

% To obtain probability $1-\alpha$, we must have 
% $$\hat{p} - Z_{\frac{\alpha}{2}} \sqrt{\frac{\hat{p}(1-\hat{p})}{k}} \leq p \leq \hat{p} + 
% Z_{\frac{\alpha}{2}} \sqrt{\frac{\hat{p}(1-\hat{p})}{k}},$$
% %\in (\hat{p}\rip Z_{\frac{\alpha}{2}}\sqrt{\frac{\hat{p}(1-\hat{p})}{k}})$,
% where $Z_{\frac{\alpha}{2}}$ is obtained from tables 
% ($\alpha = 0,1 \Rightarrow Z_{\frac{\alpha}{2}} \neq 2$; 
% $\alpha = 0,0001 \Rightarrow Z_{\frac{\alpha}{2}} \leq 4$) 
%
Next, assume that $\hat{p}$ falls into the slice $S_{l,u}$, with $l$ and $u$ its lower
and upper boundaries, respectively.
Then, as long as $\hat{p} - Z_\frac{\alpha}{2}\sqrt{\frac{\hat{p}(1-\hat{p})}{k}}
> l$ and
$\hat{p} + Z_\frac{\alpha}{2}\sqrt{\frac{\hat{p}(1-\hat{p})}{k}} \leq u$, the slice estimate 
is exact with a confidence coefficient of $100(1-\alpha)\%$.
Let $d=\min(\hat{p}-l,u-\hat{p})$, then we need
%Assume without loss of generality that $\hat{p}-l \leq u-\hat{p}$, then we need
\begin{eqnarray}
d &\geq& Z_{ \frac{\alpha}{2} } \sqrt{\frac{\hat{p}(1-\hat{p}) }{ k }}, \notag \\
k &\geq& \left( Z_\frac{\alpha}{2} \frac{ \sqrt{ \hat{p}(1-\hat{p}) } }{d}
\right)^2.\notag
\end{eqnarray}
%Consequently, we obtain:
%$$k \geq \left( Z_\frac{\alpha}{2} \frac{ \sqrt{\hat{p}(1-\hat{p})} }{ \min(\hat{p}-l, u-\hat{p})} \right)^2.\Box$$
\end{proof}

To conclude, under reasonable assumptions all node estimate its slice with confidence coefficient $100(1 - \alpha)\%$, after a finite number of message receipts. 
Moreover a node closer to the slice boundary needs more messages than a node far from the boundary.

\subsection{Simulation Results}

This section evaluates the ranking algorithm by focusing on three
different aspects.  
First, the performance of the ranking algorithm is
compared to the performance of the ordering algorithm\footnote{We 
omit comparison with JK since the performance obtained with mod-JK are either 
similar or better.} in a large-scale system where the distribution 
of attribute values does not vary over time.
Second, we investigate if sufficient uniformity is achievable in reality using a dedicated 
protocol.
%
%study the feasibility of our approach.  
%Each node
%requires to be associated with a random set of neighbors, chosen 
%at uniform. 
%We present a view management protocol that gives 
%performance results very similar to the desired 
%uniform simulation.
%
Third, the ranking algorithm and ordering algorithm are compared in 
a dynamic system where the distribution of attribute values may 
change.

%This Section evaluates the impact of dynamics on our algorithms.
%The ranking algorithm presented in Figure~\ref{alg:attr} is compared to 
%the ordering algorithm presented in the previous section.
%
For this purpose, we ran two simulations, one for each algorithms.
The system contains (initially) $10^4$ nodes and each view contains $10$ uniformly drawn random 
nodes and is updated in each cycle.
The number of slices is 100, and we present the evolution of the slice disorder measure
over time.

\begin{figure*}
  \begin{center}
    \subfigure[]
    { 
      \label{fig:comparison}    
      \resizebox{2.6in}{!}{\includegraphics[scale=0.5,angle=270,clip=true]{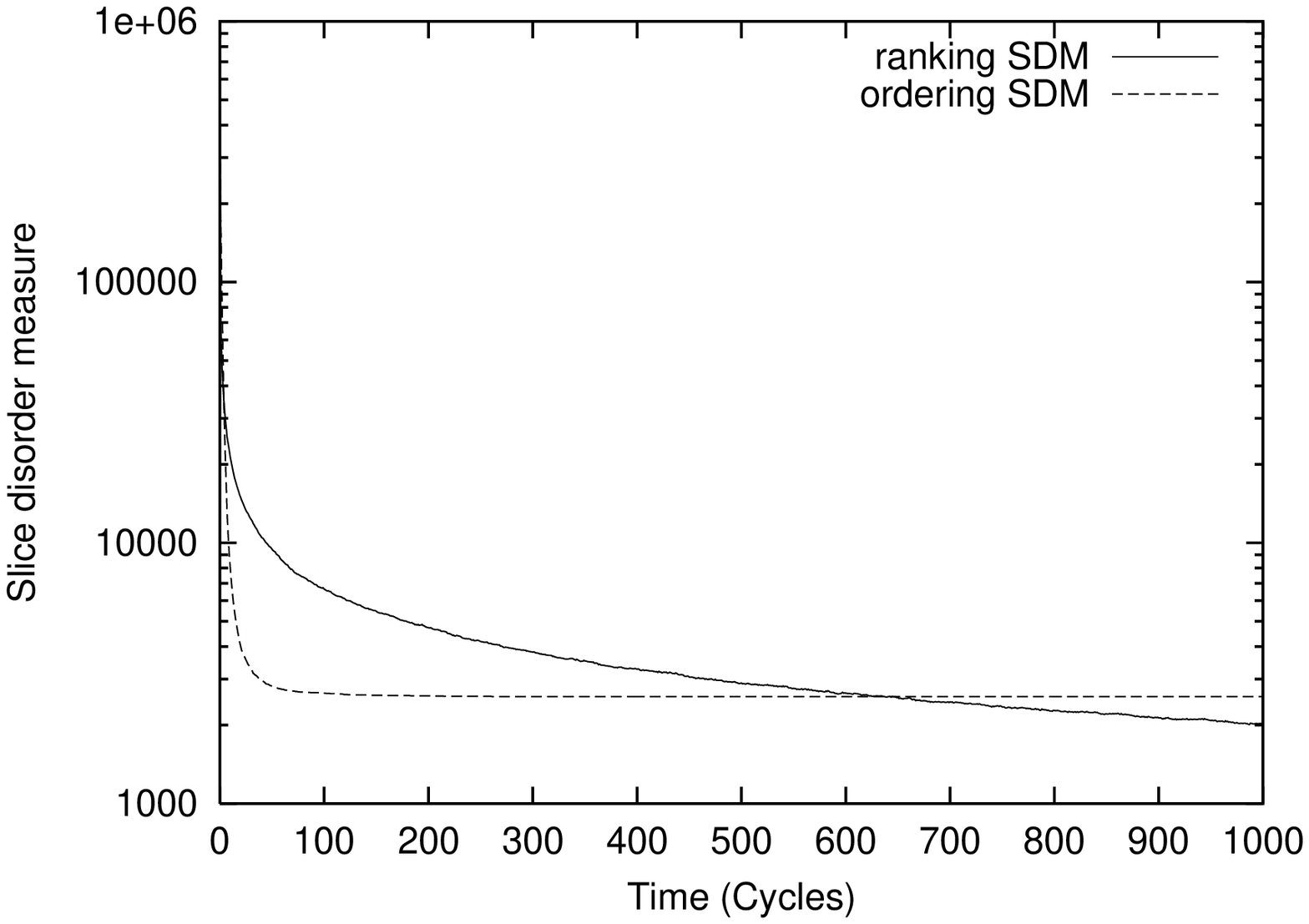}}
    }
    \subfigure[]
    { 
      \label{fig:convergence3}
      \resizebox{2.6in}{!}{\includegraphics[scale=0.5,angle=270,clip=true]{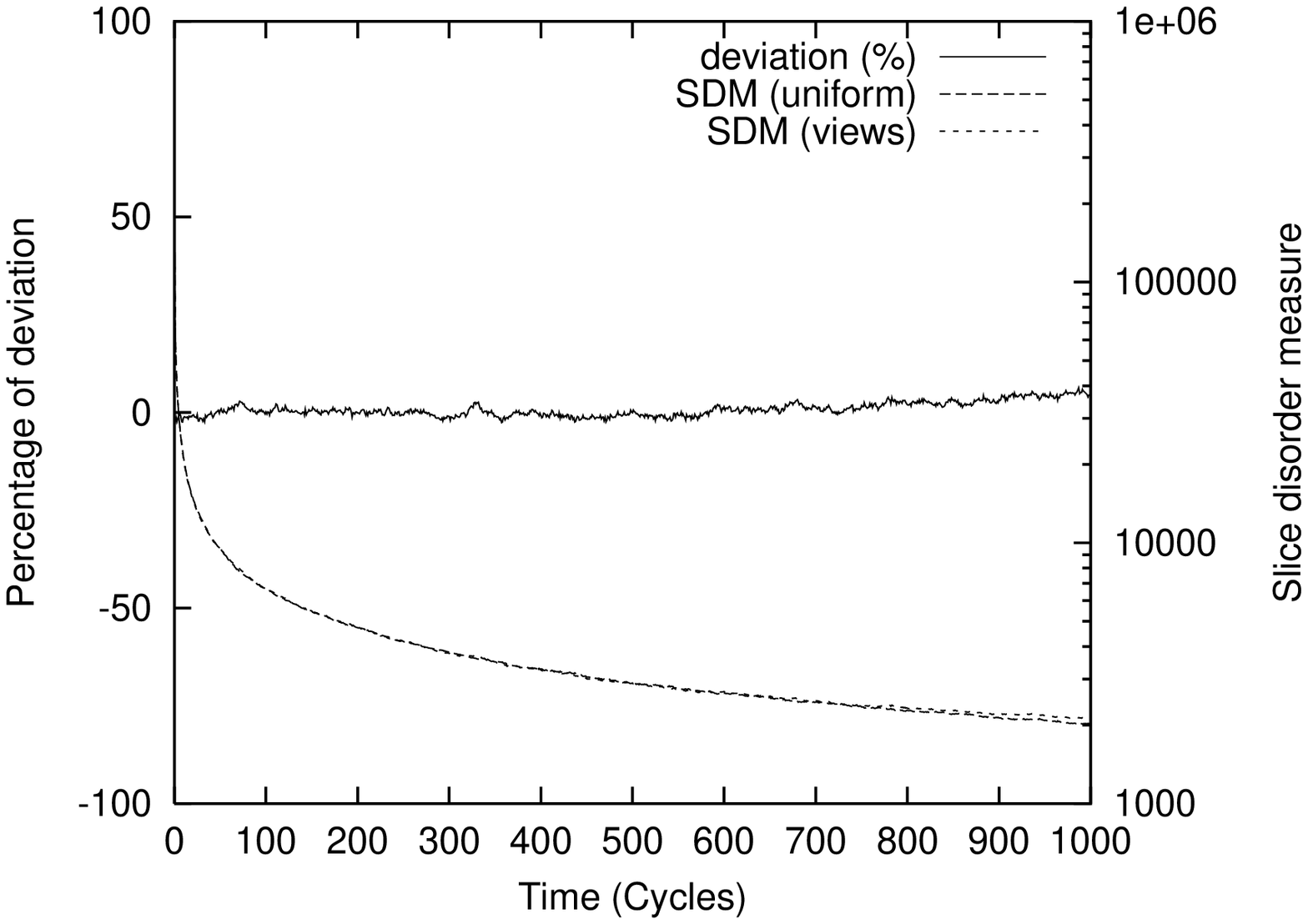}}
    }
    \subfigure[]
    { 
      \label{fig:failures2}
      \resizebox{2.6in}{!}{\includegraphics[scale=0.5,angle=270,clip=true]{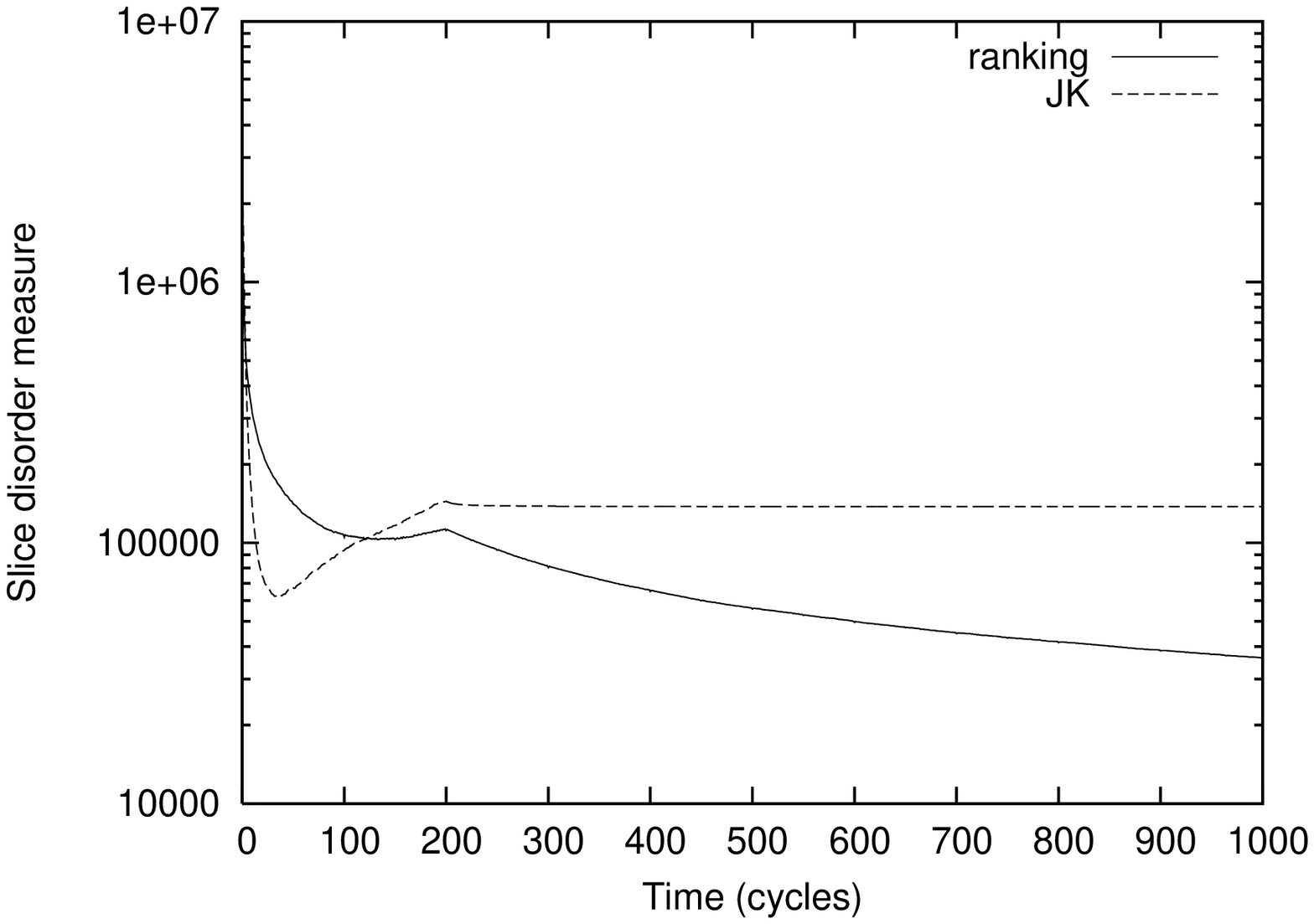}}
    }
    \subfigure[]
    {
    \label{fig:failures1}
          \resizebox{2.6in}{!}{\includegraphics[scale=0.5,angle=270,clip=true]{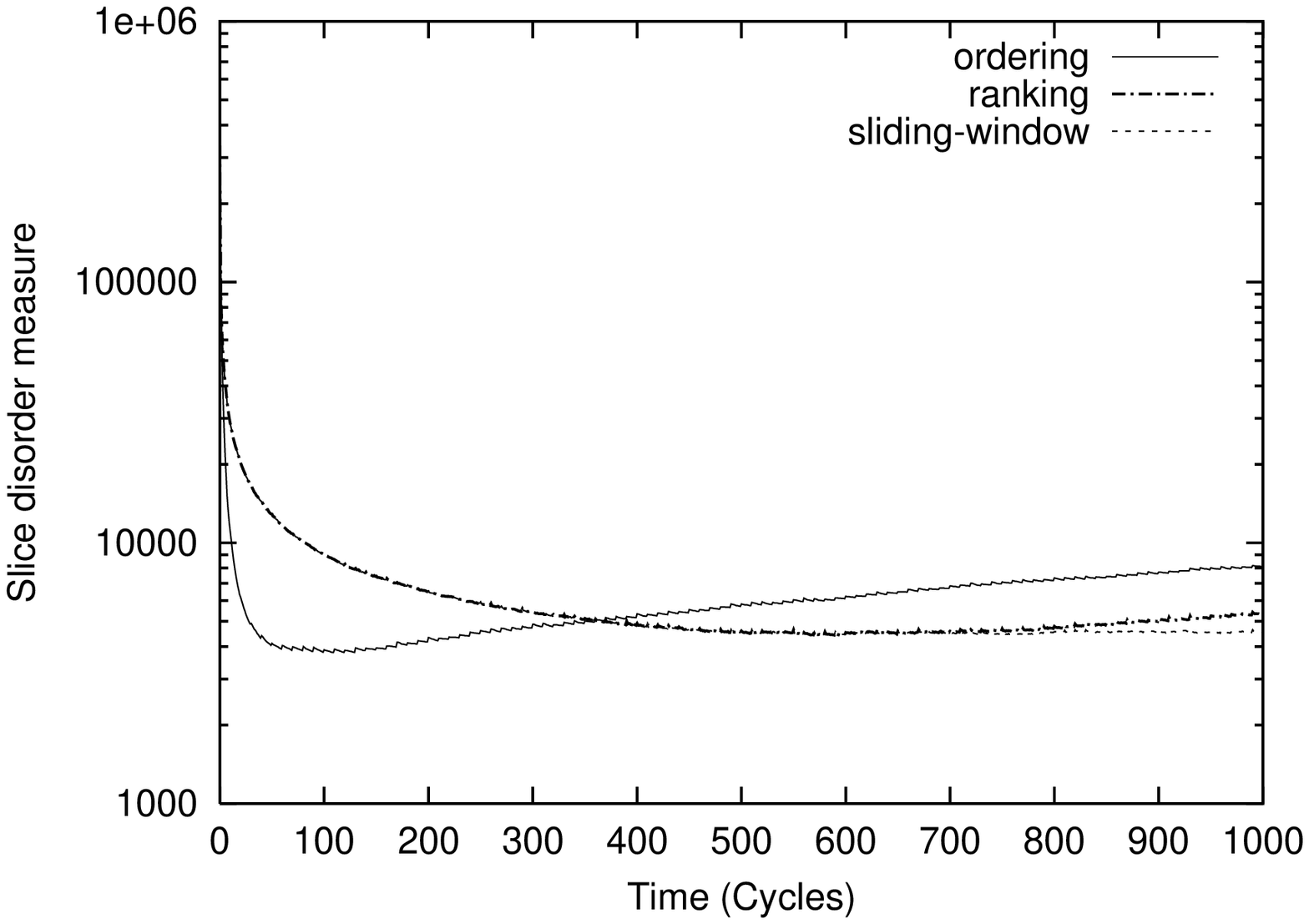}}
    }
    \caption{
       (a) Comparing performance of the ordering algorithm and the ranking algorithm (static case).
       (b) Comparing the ranking algorithm on top of a uniform drawing or a Cyclon-like protocol.
       (c) Effect of dynamics burst on the convergence of the ordering algorithm and the ranking algorithm.
       (d) Effect of a low and regular churn on the convergence of the ordering algorithm and the ranking algorithm.
       }
  \end{center}
\end{figure*}

\subsubsection{Performance Comparison in the Static Case}
Figure~\ref{fig:comparison} compares the ranking algorithm to the ordering algorithm
while the distribution of attribute values do not change over time 
(varying distribution is simulated below).

The difference between the ordering algorithm and the ranking algorithm
indicates that the ranking algorithm gives a more precise result (in terms of node to 
slice assignments) than the ordering algorithm.  
More importantly, the slice disorder measure obtained by the ordering algorithm is
lower bounded while the one of the ranking algorithm is not.
Consequently, this simulation shows that the ordering algorithm might fail in 
slicing the system while the ranking algorithm keeps improving its accuracy over 
time.

\subsubsection{Feasibility of the Ranking Algorithm}
%Figure~\ref{fig:convergence3} shows that even though the underlying membership
%protocol might impact on the convergence time of the ordering and ranking algorithms,
%it does not impact on their performance in terms of precision. 
Figure~\ref{fig:convergence3} shows that the ranking algorithm does not need artificial
uniform drawing of neighbors.  Indeed, an underlying view management protocol might 
lead to similar performance results.
%
%the ordering as the ranking algorithm
%present performance independent from the underlying.  First this shows that the result
%achieve by the ordering algorithm is limited regardless the underlying
%sampling protocol.  Second, this shows that the performance in term of 
%precision of the ranking algorithm are independent
%
%shows that our approach is realistic.  
%In the previous simulations the neighbors chosen for communication exchanges
%were selected uniformly at random among all nodes in the system.  
%Here, we simulate the ranking algorithm using an underlying view exchange 
%protocol, called Cyclon~\cite{VGS05}.
In the presented simulation we used an artificial protocol, drawing neighbors
randomly at uniform in each cycle of the algorithm execution, and the variant
of the Cyclon~\cite{VGS05} view management protocol presented above. 
Those underlying protocols are distinguished on the figure using terms 
"uniform" (for the former one) and "views" (for the later one).
As said previously, the Cyclon protocol consists of exchanging views 
between neighbors such that 
the communication graph produced shares similarities with a random graph.
This figure shows that both cases give very similar results.  
The SDM legend is on the right-handed vertical axis while
the left-handed vertical axis indicates what percentage the SDM difference
represents over the total SDM value.  At any time during the simulation 
(and for both type of algorithms) its 
value remains within plus or minus $7\%$.
The two SDM curves of the ranking algorithm almost overlap.  
Consequently, the ranking algorithm and the variant of Cyclon 
presented in subection~\ref{sec:modifiedjk} achieve very similar result.
%
%Surprisingly however, the variant of Cyclon as an underlying algorithm 
%leads seemingly to better performance than drawing the samples uniformly at random.
%In a recent work~\cite{I05} about the experimental analysis of Cyclon compare to a random graph, 
%the clustering coefficient provided by the communication graph
%of Cyclon is slightly smaller than the clustering coefficient provided by the random 
%graph. Based on this observation, it is reasonable to assume that new nodes are more 
%likely discovered with Cyclon than a random graph, and this might explain this 
%improvement.

To conclude, the variant of Cyclon algorithm presented in the previous section can be used 
with the ranking algorithm to provide the shuffling of views.
%despite this observation the similarity of the use of Cyclon and the use 
%of randomly drawing shows the feasibility of our ranking algorithms in reality.

\subsubsection{Performance Comparison in the Dynamic Case}

In Figure~\ref{fig:failures2} each of the two curves represents the 
slice disorder measure obtained over time using the ordering algorithm and 
the ranking algorithm respectively.  
We simulate the churn such that 0.1\% of nodes leave and 0.1\% of the nodes 
join in each cycle during the 200 first cycles. We observe how the SDM converges. 
The churn is reasonably and pessimistically tuned compared to recent
experimental evaluations~\cite{SR06} of the session duration in three well-known P2P 
systems.\footnote{In~\cite{SR06}, roughly all nodes have left the system after 1 day while there are still 
50\% of nodes after 25 minutes. In our case, assuming that in average a cycle lasts 
one second would lead to more than 54\% of leave in 9 minutes.}

The distribution of the churn is correlated to the attribute value of the nodes. 
The leaving nodes are the 
nodes with the lowest attribute values while the entering nodes have higher attribute 
values than all nodes already in the system.  The parameter choices are motivated by 
the need of simulating a system in which the attribute value corresponds to the 
session duration of nodes, for example.

The churn introduces a significant disorder in the system which 
counters the fast decrease.  When, the churn stops, the ranking algorithm readapts 
well the slice assignments: the SDM starts decreasing again.  However, 
in the ordering algorithm, the convergence of SDM gets stuck. 
This leads to a poor slice assignment accuracy.

In Figure~\ref{fig:failures1}, 
each of the two curves represent the slice disorder measure obtained over time
using the ordering algorithm, the ranking algorithm, and a modified version of the ranking
algorithm using attribute values recorded in a sliding-window, respectively.
(The simulation obtained using sliding windows is described in the next subsection.)
The churn is diminished and made more regular than in the previous simulation such that
0.1\% of nodes leave and 0.1\% of nodes join every 10 cycles.
%Recall that the slice
%disorder measure (SDM) represents the sum for all nodes of the distance between the 
%slice the node thinks it belongs to and the slice it really belongs to.

The curves fits a fast decrease (superlinear in the number of cycles) at the 
beginning of the simulation.
At first cycles, the ordering gain is significant making the impact of churn
negligible.  This phenomenon is due to the fact that SDM decreases rapidly 
when the system is fully disordered.  Later on, however, the decrease slope 
diminishes and the churn effect reduces the amount of nodes with a low attribute 
value while increasing the amount of nodes with a large attribute value.
This unbalance leads to a messy slice assignment, that is, each node must quickly
find its new slice to prevent the SDM from increasing.
In the ordering algorithm the SDM starts increasing from cycle 120. Conversely, 
with the ranking algorithm the SDM starts increasing not earlier than at cycle 730.
Moreover the increase slope is much larger in the former algorithm than in the latter 
one.

Even though the performance of the ranking algorithm are really significant, its 
adaptiveness to churn is not surprising.  Unlike the ordering algorithm, the 
ranking one keeps re-estimating the rank of each node depending on the 
attribute values present in the system.
Since the churn increases the attribute values present in the 
system, nodes tend to receive more messages with higher attribute values and 
less messages with lower attribute values, which turns out to keep the SDM 
low, despite churn. Further on, we propose a solution  based on 
sliding-window technique to limit the increase of the SDM in the ranking 
algorithm.

To conclude, the results show that when the churn is related to the attribute 
(e.g., attribute represents the session duration, uptime of a node), 
then the ranking algorithm is better suited than the ordering algorithm.

\subsubsection{Sliding-window for Limiting the SDM Increase}

In Figure~\ref{fig:failures1}, the "sliding-window" curve 
presents a slightly modified version of the ranking algorithm
that encompasses SDM increase due to churn correlated to attribute 
values.  Here, we present this enrichment.

In Section~\ref{sec:ranking}, the ranking algorithm specifies that
each node takes into account all received messages.
More precisely, upon reception of a new message each node $i$ re-computes
immediately its rank estimate and the slice it thinks it belongs to 
without remembering the attribute values it has seen.
Consequently the messages
received long-time ago have as much importance as the fresh messages
in the estimate of $i$.
The drawback, as it appeared in Figure~\ref{fig:failures1} of Section~\ref{sec:simu2}, is that if  
the attribute values are correlated to churn, then the precision of the algorithm might diminish.

To cope with this issue, the previous algorithm can be easily enriched
in the following way.  Upon reception of a message, each node $i$
records an information about the attribute value received in a fixed-size ordered set 
of values.  Say this set is a first-in first-out buffer such that
only the most recent values remain.
Right after having recorded this information, node $i$ can re-compute 
its rank estimate and its slice estimate based on the most relevant 
piece of information (having discarded the irrelevant piece). 
Consequently, the estimate would rely only on fresh attribute values 
encountered so that the algorithm would be more tolerant to changes 
(e.g., dynamics or non-uniform evolution of attribute values).
Of course, since the analysis (cf. Section~\ref{sec:analysis2}) shows that nodes close to 
the slice boundary 
require a large number of attribute values for estimating precisely their 
estimates, it would be unaffordable to record all these last attribute 
values encountered due to space limitation.

Actually, the only necessary relevant information of a message
is simply whether it contains a lower attribute value than the attribute
value of $i$, or not.  Consequently, a single bit per message would be sufficient 
to record the necessary information (e.g., adding a 1 meaning that the attribute value
is lower, and 0 otherwise).  Thus, even though a node $i$ would require 
$10^4$ messages to rightly estimate its slice (with high probability), 
node $i$ simply needs to allocate an array of size 
%$\lceil 10^4/(8*1024)\rceil =  1,23$ KB.
$10^4/(8*1000) =  1,25$ kB.

As expected, Figure~\ref{fig:failures1} shows that the sliding-window method
applied to the ranking algorithm prevents its SDM from increasing.  Consequently, at some
point in time, the resulting slice assignment may become even more accurate.

\section{Conclusion}
\label{conclusion}

\subsection{Summary}
Peer to peer systems may now be turned into general frameworks on top of which 
several applications might cohabit. To this end, allocating resources to applications,
according to their needs require specific algorithms to partition the network in a relevant
way. The ordered slicing algorithm proposed in \cite{JK06} provided a first attempt to 
``slice'' the network, taking into account the potential heterogeneity of nodes. This algorithm
relies on each node drawing a random value uniformly and swapping continuously those random values,
 with candidate
nodes, so that the order between attributes values (reflecting the capabilities of nodes) and random ones
match. Results from \cite{JK06} have shown that slices can be maintained efficiently and 
in large-scale systems even in the presence of churn. 

In this report, we first proposed an improvement over the initial sorting algorithm based on a judicious
choice of candidate nodes to swap values. This is based on each node being able to estimate locally
the potential decrease of the global disorder measure. We provided an analysis along with some simulation
results showing that the convergence speed is significantly improved.
We then identified two issues related to the use of static random values. 
The first one refers to the fact that slice assignment heavily depends on 
the degree of uniformity of the initial random value. 

The second is related to the fact that
once sorted along one attribute axis, the churn (or failures) might be correlated to the
attribute, therefore leading to a unrecoverable skewed distribution of the 
random values resulting in a wrong slice assignment.
Our second contribution is an algorithm enabling nodes to continuously
re-estimate their rank relatively to other nodes based on their sampling of
the network. 

\subsection{Perspective}
%\subsection{Cyclon as an Underlying Sampling Algorithm}
%This report presents two algorithms relying on a variant of Cyclon to benefit from
%the distribution quasi-uniform of the neighbors.
%The simulation with the variant of Cyclon shows that the two algorithms 
%are independent from the underlying physical overlay.
%More technically, this shows the feasibility of our approach by providing 
%the distribution requirement of the samples.
%
%Nevertheless, previous simulation surveys have shown that Cyclon might suffer
%from a lack of robustness in face of massive departures compared
%to other gossip-based approach such as Newscast.  This report did not
%simulate the behavior of Cyclon under such circumstances.  Despite the 
%interest of simulating the ranking algorithm on top of Cyclon and in 
%face of massive failures, such extensive experiments were out of the 
%scope of this report and could be an interesting future work.

This report used a variant of the Cyclon protocol to obtain quasi-uniform
distribution of neighbors.  There are various protocols that might be used 
for different purpose.  For instance, Newscast can be used for its resilience
to very high dynamics as in~\cite{JK06}.  Some other protocols exist 
in the literature.  Deciding exactly how to parameterize the underlying 
peer sampling service might be an interesting future direction.

\subsection*{Acknowledgment}
We wish especially to thank M\'ark Jelasity for the fruitful discussions we 
had and the time he spent improving this report. 
We are also thankful to Spyros Voulgaris for having kindly shared his work 
on the Cyclon development. The work of A.  Fern\'andez and E. Jim\'enez was partially 
supported by the Spanish MEC under grants TIN2005-09198-C02-01, 
TIN2004-07474-C02-02, and TIN2004-07474-C02-01, and the 
Comunidad de Madrid under grant S-0505/TIC/0285. The work of A.  Fern\'andez was done 
while on leave at IRISA, supported by the Spanish MEC under grant 
PR-2006-0193.

\bibliographystyle{plain}
\bibliography{slice}

\end{document}